\documentstyle[aps,psfig,epsfig]{revtex}

\newcommand{\pr}{\prime}
\newcommand{\na}{\nabla}

\newcommand{\ep}{\epsilon}

\newcommand{\lraw}{\longrightarrow}

\newcommand{\pa}{\partial}

\newcommand{\td}{\tilde}
\newcommand{\sla}[1]{\slash\!\!\! #1}
\newcommand{\sct}[1]{\over{|\hspace{#1 in}|}}

\begin{document}

 \draft
\title{Resummation Studies on Vector Meson Decays and Chiral Symmetry
Spontaneously Breaking in Chiral Constituent Quark Model}
\author{Yi-Bin Huang\footnote{E-mail address:
huangyb@mail.ustc.edu.cn}}
\address{Center for Fundamental Physics,
University of Science and Technology of China\\
Hefei, Anhui 230026, P. R. China}
\author{Xiao-Jun Wang\footnote{E-mail address: wangxj@itp.ac.cn}}
\address{
Institute of Theoretical Physics, Beijing, 100080, P. R. China}
\author{Mu-Lin Yan\footnote{E-mail address: mlyan@staff.ustc.edu.cn}}
\address{CCST(World Lab), P. O. Box 8730, Beijing, 100080, P. R. China \\
  and\\
 Center for Fundamental Physics,
University of Science and Technology of China\\
Hefei, Anhui 230026, P. R. China\footnote{mail address}}
\date{\today}
\maketitle

\begin{abstract}
We study on-shell decays of light vector meson resonances $\rho$,
$K^*$ and $\phi$ in the framework of chiral constituent quark
model using resummation calculations. Such studies are necessary
for showing that chiral dynamics works well at this energy scale.
The effective action is derived by proper vertex method, where
resummation of all orders of momentum expansion is accomplished.
Also studied are the loop effects of pseudoscalar meson, which
play an important role at this energy scale. The numerical results
agree well with the experimental data. A new method to explore the
chiral symmetry spontaneously breaking (CSSB) is proposed. It is
found that the unitarity of the effective meson theory resulted
from resummation derivations demands an upper-limit to the
momentum of vector meson. This upper-limit, being critical point,
is just the energy scale of CSSB, and is found to be
flavor-dependent.
\end{abstract}
\pacs{12.39.-x,11.30.Rd,11.30.Qc,12.40.Yx,12.40.Vv,13.25.-k,13.75.Lb}

\section{Introduction}

The Chiral Symmetry Spontaneously Breaking (CSSB) is an important
feature in the theory of QCD, which governs the dynamics of
hadrons at very low energies. It is generally believed that there
is an energy scale $\Lambda_{\rm CSSB}\simeq 2\pi F_\pi \simeq
1.2$GeV, below which the chiral symmetry $SU(3)_L\times SU(3)_R$
is spontaneously breaking to $SU(3)_{L+R}$ associated with eight
Goldstone bosons\cite{Wein79,Coleman,MG84}. Under this strong
symmetry constraint, effective Lagrangian of mesons, ${\cal
L}_{\rm eff}$, can be constructed with numbers of
parameters\cite{Wein79,GL85a}. This is the base of the chiral
perturbation theory (ChPT)\cite{GL85a}, and hence ChPT can be
thought of as a rigorous QCD theory.  As the typical energy scale
$p$ for corresponding hadron dynamics is much less than
$\Lambda_{\rm CSSB}$, i.e., $p/\Lambda_{\rm CSSB}<<1$, it serves
as a good approximation to take a few leading- and next to the
leading terms in ${\cal L}_{\rm eff}$ in $p-$(or in momentum, or
in derivative) expansion . Thus, the ChPT calculations become to
be practicable. Usually, in practice, the calculations in ChPT are
up to $O(p^4)$. The predictions of such $O(p^4)$-calculations
describe pseudoscalar meson physics at very low energies quite
well.

However, as the energies go up to the meson resonance region (say
$E \sim M_\rho$), two novel ingredients should be taken into
account: 1, the meson resonances will be excited and they should
emerge in the theory; 2, since the energies go up, the
$p/\Lambda_{\rm CSSB}$ will be not much less than $1$. For
$\rho$-decays,  $p\sim 0.77$GeV and $p/\Lambda_{\rm CSSB} \sim
0.64$; for $K^*$, $p\sim 0.892$GeV  and $p/\Lambda_{\rm CSSB} \sim
0.74$; for $\phi$, $p\sim 1.02$GeV  and $p/\Lambda_{\rm CSSB} \sim
0.85$. In these cases, the calculations based on taking only a few
leading terms in $p-$expansion (i.e., so called $O(p^4)-$
calculations) will no longer be legitimate. In order to
investigate the vector meson decays, all terms of the
$p-$expansions should be taken into account, and then be summed
over. We will call it resummation studies on the corresponding
vector meson processes. Obviously, for vector meson resonance
physics at low energies, such resummation studies  are necessary
and meaningful even though it may be a heavy work. If such studies
based on chiral expansion can be performed and the results are
reasonable well, then one can conclude that the chiral dynamics
(or chiral effective Lagrangian theory) works up to this vector
meson energy scale. If not, we will have no reason to think so.

Another motive in this paper is to explore the CSSB of QCD in a
non-trivial and realistic QCD-inspired model. CSSB is a prior
hypothesis in ChPT. Hence, the success of ChPT provides an
indirect evidence for existence of CSSB in QCD. The mechanism of
CSSB has been widely discussed in the
literature\cite{Nambu}\cite{Roberts}. However, that how to prove
it and how to derive and then to determine the critical energy
scale $\Lambda_{\rm CSSB}$ from the fundamental QCD theory still
remain to be settled\cite{Leutwyler}. Therefore, it is still
interesting and meaningful to study this subject in more realistic
models and in new non-perturbative methods. CSSB in QCD could be
thought of as a kind of quantum phase transition phenomenon in the
quantum field theories, which is caused by quantum fluctuations in
the system\cite{Sachdev}. It is well known that as $p$ below
$\Lambda_{\rm CSSB}$ (or after CSSB), the quantum dynamical
freedoms are meson fields and the dynamics is described in chiral
effective meson Lagrangian. The $S$-matrices of this Lagrangian
field theory have to be unitary, which belongs to the first
principle requirement in quantum theories. Thus, as one had a
chiral effective meson Lagrangian with all order-terms in $p$- (or
space-time derivative-) expansions, the following question can be
asked: In what range of $\sqrt{p^2}$ the $S$-matrices yielded by
the Feynman rules of the theory are unitary? The answer will lead
to the determination of $\Lambda_{\rm CSSB}$ because the
upper-limit of this $p$-range should just be $\Lambda_{\rm CSSB}$.
In other words, as above this $p$-upper-limit, i.e., $\Lambda_{\rm
CSSB}$, the quantum field description of this chiral effective
meson Lagrangian system will collapse. This is precisely a
critical phenomenon. In conception of Heisenberg's uncertainty
principle, the quantum fluctuations of the system in the
coordinate space are arisen from its momentum $p$: larger distance
physics associating with relatively smaller quantum fluctuations
corresponds to smaller $p$, and smaller distance one with larger
fluctuations corresponds to larger $p$. Thus, for a quantum field
system, it may transfer from order phase to disorder phase along
with $p$-increasing, and then the critical energy scale emerges in
the description of the dynamics. It is meaningful and interesting
to reveal this scale by applying non-perturbation method to a
quantum field system and by examining the unitarity of the theory.
In this paper we shall try to realize this idea, i.e., we shall
use the resummation derivation method to explore the unitarity of
the chiral constituent quark model with vector mesons (see below),
and then to determine the critical scale $\Lambda_{\rm CSSB}$.

In the literature, there are several schemes to extend the chiral
symmetry considerations to be including vector meson resonances,
and then the corresponding ChPT-like effective theories with $0^-$
and $1^\pm$ mesons can be constructed and
studied\cite{Eck89}\cite{Bando}. Because there are huge number of
unknown-parameters in high order terms of $p-$expansion in this
kind of theories, it is impracticable to perform resummation
studies in the formalism of ChPT-like theories with vector mesons.
Actually, most of all calculations in the literature in these
ChPT-like theories are limited to be of $O(p^4)$ or ${\cal
O}(p^6)$ and to be of the leading order of $1/N_c$-expansion
\cite{Op46}. This situation, of course, is not satisfactory for
the studies of vector meson physics even though the theories seem
to be model-independent. In this paper we try to provide a
phenomenon study based on systemical resummation calculations to
processes of $\rho \rightarrow \pi\pi$, $\rho^0 \rightarrow
e^+e^-$, $K^* \rightarrow K\pi$, $\phi \rightarrow K^+K^-$ and
$\phi \rightarrow K^0_LK^0_S$ in a realistic QCD-inspired model.

As a QCD-inspired model, the Chiral Quark Model (ChQM) (or
Numbu-Jona-Lasinio  version models and its extensions) has been
extensively studied in hadron
physics\cite{MG84,Nambu,Chan85,Esp90,ENJL,Li95,Wang98,WY00,Wangrho,Wangvect}.
The starting point of the model is a chiral constituent quark
lagrangian with dynamical Goldstone bosons\cite{MG84}. The spin-1
mesons are included into the model by using the WCCWZ
realization\cite{Wein68,WCCWZ}. In
refs.\cite{WY00,Wangrho,Wangvect}, two of us have investigated
this model in $p$-resummation manner for $\rho$-decays. In this
paper we shall recapture the resummation studies in
refs.\cite{WY00,Wangrho,Wangvect} and make it more precise, and
then extend it to $K^*$-, $\phi$-decay processes. Furthermore, we
propose a new method to determine the $\Lambda_{\rm CSSB}$. We
shall use large-$N_c$ expansion and optic theorem to prove a
necessary  condition for the unitarity of the theory, which has to
be satisfied by meson's transition amplitudes. Then we use the
Feynman rules to calculate the transition amplitudes of vector
meson decays, and compare the results with the requirement of the
necessary condition of the unitarity, and then the $\Lambda_{\rm
CSSB}$ is determined. This determination is regularization scheme
free.

Specifically, the follows will be shown in this paper: 1, In order
to perform the $p$-resummation derivation to the effective meson
Lagrangian described vector meson decays, a method called as
proper vertex expansion\cite{WY00,Wangrho,Wangvect} (rather than
the Schwenger proper time method\cite{Sch54,Ball89}) is used to
calculate the quark loop contributions to it. It is shown that the
power series of momentum expansion for the vector meson decay
amplitudes converge slowly. This fact indicates that the
$p$-resummations are necessary indeed for the vector meson decays;
2, Since both constituent quarks and the Goldstone bosons are
dynamical freedom fields in the ChQM, in the calculations for
getting the effective meson Lagrangian at one loop level both
contributions due to the quark loop and ones due to the Goldstone
boson loop have to be taken into account. Considering the
contributions of quark loops and ones of Goldstone boson loops are
of $O(N_c)$ and $O(1)$ in $1/N_c$-expansion respectively,
consequently, any consistent loop-expansion calculations in the
ChQM must include the contributions from the next to leading order
in $1/N_c$-expansion in the model. In this paper, we shall
calculate both quark loops and Goldstone boson loops for
$V\rightarrow \Phi\Phi$ ($V$ and $\Phi$ are vector- and
pseudoscalar mesons respectively) in ChQM. The analytical
calculations to the corrections of the next to the leading order
of ${\cal O}(1/N_c)$-expansion are somehow heavy, but it is
necessary; 3,The parameters in the effective meson lagrangian
derived from the above procedure can been fixed by meeting the
requirements of KSRF sum rule\cite{KSRF}, Zweig rule forbidden to
$\phi \rightarrow \pi\pi$, beta decay of neutron and by matching
the low energy limit of this theory with the constraints of ChPT;
4, The low energy limit of the effective meson field theory of
ChQM is checked and it is shown that the results are consistent
with ChPT, and hence ChQM is of a legitimate QCD-inspired model at
very low energies (see Appendix A); 5, The decay widths for $\rho
\rightarrow \pi\pi$, $\rho^0 \rightarrow e^+e^-$, $K^* \rightarrow
K\pi$, $\phi \rightarrow K^+K^-$ and $\phi \rightarrow K^0_LK^0_S$
are calculated in this parameter-free theory and the predictions
are compared with data; 6, Based on the results of the resummation
studies, we derive the $\Lambda_{CSSB}$ and it is found out that
$\Lambda_{CSSB}$ is flavor-dependent.

The contents of this paper are organized as following: In section
II we introduce the model with giving the notations; Section III
is devoted to illustrate the proper vertex expansion; In section
IV, the kinetic terms of vector mesons are derived; Section V, the
quark loop contributions to vector meson decays;  Section VI, the
Goldstone boson loop contributions to vector meson decays; Section
VII, the numerical results; Section VIII, unitarity and large
$N_c$ expansion. A necessary condition for the unitarity of the
meson theory deduced from ChQM is revealed in this section;
Section IX, determination of $\Lambda_{\rm CSSB}$: i.e., applying
the necessary condition of the unitarity, the upper limit of $p$
is derived, and then $\Lambda_{\rm CSSB}$ is determined. Finally,
we provide a brief summary and discussion. In the Appendices, we
provide derivations of the low-energy limit of the theory, and
show how to perform parametrization of the quadratic divergence
emerged in the meson loop calculations in the text. The paper is
self-consistent.

\section{the model}

For understanding the hadron physics below CSSB scale, Manohar and
Georgi provides a QCD-inspired description on the simple
constituent quark model \cite{MG84} (call it as simple-ChQM
hereafter). At chiral limit, it is parameterized by the following
$SU(3)_{V}$ invariant chiral constituent quark Lagrangian
\begin{eqnarray}\label{2.1}
{\cal L}_{\chi}&=&i\bar{q}(\sla{\pa}+\sla{\Gamma}+
  g_{A}{\slash\!\!\!\!\Delta}\gamma_5)q-m\bar{q}q
   +\frac{F^2}{16}<\nabla_\mu U\nabla^\mu U^{\dag}>.
\end{eqnarray}
Here $<\cdots>$ denotes trace in SU(3) flavor space,
$\bar{q}=(\bar{u},\bar{d},\bar{s})$ are constituent quark fields,
$g_{A}=0.75$ is fitted by beta decay of neutron. The $\Delta_\mu$
and $\Gamma_\mu$ are defined as follows,
\begin{eqnarray}\label{2.2}
\Delta_\mu&=&\frac{1}{2}[\xi^{\dag}(\pa_\mu-ir_\mu)\xi
          -\xi(\pa_\mu-il_\mu)\xi^{\dag}], \nonumber \\
\Gamma_\mu&=&\frac{1}{2}[\xi^{\dag}(\pa_\mu-ir_\mu)\xi
          +\xi(\pa_\mu-il_\mu)\xi^{\dag}],
\end{eqnarray}
and covariant derivative are defined as follows
\begin{eqnarray}\label{2.3}
\nabla_\mu U&=&\pa_\mu U-ir_\mu U+iUl_\mu=2\xi\Delta_\mu\xi,
  \nonumber \\
\nabla_\mu U^{\dag}&=&\pa_\mu U^{\dag}-il_\mu U^{\dag}+iU^{\dag}r_\mu
  =-2\xi^{\dag}\Delta_\mu\xi^{\dag},
\end{eqnarray}
where $l_\mu=v_\mu+a_\mu$ and $r_\mu=v_\mu-a_\mu$ are linear
combinations of external vector field $v_\mu$ and axial-vector
field $a_\mu$, $\xi$ associates with non-linear realization of
spontaneously broken global chiral symmetry $G=SU(3)_L\times
SU(3)_R$ introduced by Weinberg \cite{Wein68},
\begin{equation}\label{2.4}
\xi(\Phi)\rightarrow
g_R\xi(\Phi)h^{\dag}(\Phi)=h(\Phi)\xi(\Phi)g_L^{\dag},\hspace{0.5in}
 g_L, g_R\in G,\;\;h(\Phi)\in H=SU(3)_{V}.
\end{equation}
Explicit form of $\xi(\Phi)$ is usually taken as
\begin{equation}\label{2.5}
\xi(\Phi)=\exp{\{i\lambda^a \Phi^a(x)/2\}},\hspace{1in}
U(\Phi)=\xi^2(\Phi),
\end{equation}
where $\lambda^1,\cdots,\lambda^8$ are SU(3) Gell-Mann
matrices in flavor space, and the
Goldstone bosons $\Phi^a$ are treated as pseudoscalar meson
octet:
\begin{equation}\label{2.6}
\Phi(x)=\lambda^a \Phi^a(x)=\sqrt{2}
\left(\begin{array}{ccc}
       \frac{\pi^0}{\sqrt{2}}+\frac{\eta_8}{\sqrt{6}}
            &\pi^+ &K^+   \\
    \pi^-&-\frac{\pi^0}{\sqrt{2}}+\frac{\eta_8}{\sqrt{6}}
            &K^0   \\
       K^-&\bar{K}^0&-\frac{2}{\sqrt{6}}\eta_8
       \end{array} \right).
\end{equation}
 The transformation law under SU(3)$_{V}$ are
\begin{equation}\label{2.7}
  \psi\lraw h(\Phi)\psi, \hspace{0.6in}
\Delta_\mu\lraw h(\Phi)\Delta_\mu h^{\dag}(\Phi), \hspace{0.6in}
\Gamma_\mu\lraw h(\Phi)\Gamma_\mu h^{\dag}(\Phi)+h(\Phi)\pa_\mu
  h^{\dag}(\Phi).
\end{equation}
Thus the Lagrangian (~\ref{2.1}) is invariant under $G_{\rm
global}\times G_{\rm local}$. There is only one parameter in this
simple model, i.e., constituent quark mass $m$. With appropriate
choice of $m$-value, the coefficients in ChPT,
$L_1,\;\;L_2\;\;L_3\;\;L_9\;\;L_{10}$, have been derived in
refs\cite{Esp90,Wang98}. The results shown that the simple-ChQM is
consistent with ChPT. Therefore, it is substantial to take the
formulation of ChQM as our stating point.

For our purposes, the simple-ChQM must be extended to include
lowest vector meson resonances and go beyond the chiral limit. The
mass difference of constituent quarks with different flavors is
assumed to be caused by current quark masses. The light quark mass
matrix ${\cal M}={\rm diag}\{m_u,m_d,m_s\}$ is usually included in
external spin-0 fields, i.e., $\td{\chi}=s+ip$, where $s=s_{\rm
ext}+{\cal M}$, $s_{\rm ext}$ and $p$ are scalar and pseudoscalar
external fields respectively. The chiral transformation for
$\td{\chi}$ is $ \td{\chi}\rightarrow g_{_R}\td{\chi}g_{_L}^{\dag}
$. Thus $\td{\chi}$ and $\td{\chi}^{\dag}$ together with $\xi$ and
$\xi^{\dag}$ can form SU(3)$_{V}$ invariant quantities
\begin{eqnarray}\label{2.8}
S=\frac{1}{2}(\xi^{\dag}\td{\chi}\xi^{\dag}+\xi\td{\chi}^{\dag}\xi),
\hspace{1in}
P=\frac{1}{2}(\xi^{\dag}\td{\chi}\xi^{\dag}-\xi\td{\chi}^{\dag}\xi),
\end{eqnarray}
which are scalar and pseudoscalar respectively.
Then the current-quark-mass-dependent term is written
\begin{equation}\label{2.9}
-\bar{q}Sq-\kappa\bar{q}P\gamma_5q,
\end{equation}
which goes back to standard quark mass term of QCD Lagrangian,
$-\bar{\psi}{\cal M}\psi$ ($\psi$ is the corresponding current
quark fields), before CSSB at high energy for arbitrary $\kappa$.
It means that the symmetry and some underlying constrains of QCD
can not fix the couplings between pseudoscalar mesons and
constituent quarks. Hence $\kappa$ is treated as an initial
parameter of the model and will be fitted phenomenologically.

From the viewpoint of chiral symmetry only, an alternative scheme for
incorporating vector mesons was suggested by Weinberg \cite{Wein68} and
developed by Callan, Coleman et al \cite{WCCWZ}. In this treatment, vector
meson resonances $V_\mu$ transform homogeneously under SU(3)$_{V}$,
\begin{eqnarray}\label{2.10}
V_\mu\rightarrow h(\Phi)V_\mu h^{\dag}(\Phi),
\end{eqnarray}
where
\begin{equation}\label{2.11}
   V_\mu(x)={\bf \lambda \cdot V}_\mu+\lambda^0V_{\mu}^0 =\sqrt{2}
\left(\begin{array}{ccc}
       \frac{\rho^0_\mu}{\sqrt{2}}+\frac{\omega_\mu}{\sqrt{2}}
            &\rho^+_\mu &K^{*+}_\mu   \\
    \rho^-_\mu&-\frac{\rho^0_\mu}{\sqrt{2}}+\frac{\omega_\mu}{\sqrt{2}}
            &K^{*0}_\mu   \\
       K^{*-}_\mu&\bar{K}^{*0}_\mu&\phi_\mu
       \end{array} \right),
\end{equation}
and $\lambda^0=\sqrt{\frac{2}{3}}$. Then the simple-ChQM is
extended to a chiral quark model including both pseudoscalar
mesons and the lowest meson resonances, which will be called ChQM
simply hereafter. ChQM is parameterized by the following
SU(3)$_{V}$ invariant Lagrangian
\begin{eqnarray}\label{2.12}
{\cal L}_{\chi}&=&i\bar{q}(\sla{\pa}+\sla{\Gamma}+
  g_{A}{\slash\!\!\!\!\Delta}\gamma_5-i\sla{V})q-m\bar{q}q
-\bar{q}Sq-\kappa\bar{q}P\gamma_5q +\frac{F^2}{16}<\nabla_\mu
U\nabla^\mu U^{\dag}>
   +\frac{1}{4}m_0^2<V_\mu V^{\mu}>.
\end{eqnarray}
We can see that there are five initial parameters
$g_{A},\;m,\;\kappa,\;F$ and $m_0$ in ChQM ($F$ will be
renormalized). These parameters can not be determined by symmetry
but can only be fitted by experiment.

\section{proper vertex expansion}
\setcounter{equation}{0}

In chiral quark model, low energy effective action of light
hadrons is generated through loop effects of constituent quarks.
The usual way to obtain the effective action is in path integral.
Integrating out degrees of freedom of quarks, we obtain a
determinant and then regularize it by Schwinger proper time method
\cite{Sch54} or heat kernel method \cite{Ball89}. In this
framework, the effective action is expanded in powers of momentum
of mesons, and the calculations to $O(p^4)$ are
practicable\cite{Li95}. However, it is very difficult to calculate
more higher order contributions of momentum expansion. Actually,
it is impracticable. Instead of it, we shall derive the effective
action by computing the loop effects of constituent quarks
directly. In this way, the calculations are expansions of loops,
or of proper vertices of external fields rather than momentum
expansions. We call this method as proper vertex expansion
following refs.\cite{Wangrho,Wangvect}, in which all terms in
$p$-expansion are catched for concrete processes, and hence the
corresponding calculations are of resummation of $p$-expansion.

The quark part of Lagrangian~(\ref{2.12}) can be divided into two
parts:
\begin{eqnarray}\label{3.1}
{\cal L}_\chi^q&=&{\cal L}_\chi^0+{\cal L}_\chi^I,\nonumber \\
{\cal L}_\chi^0&=&\bar{q}(i\sla{\pa}-m-{\cal M})q,\nonumber \\
{\cal L}_\chi^I&=&\bar{q}\sla{V}q+\bar{q}(i\sla{\Gamma}-S')q+
\bar{q}(ig_A{\slash\!\!\!\!\Delta}-\kappa P)\gamma_5q,
\end{eqnarray}
where $S'=S-{\cal M}$.

As a consequence of the free part ${\cal L}_\chi^0$,
 the propagators of constituent quarks are
\begin{eqnarray}\label{3.2}
\stackrel{\sct{0.3}}{q_i^a (x)\bar{q}_j^b}(y)=\delta_{ab}\int
\frac{d^4p}{(2\pi)^4}e^{-ip\cdot (x-y)}\frac{i(\sla{p}+M_a)_{ij}}
{p^2-M_a^2+i\epsilon}=\delta_{ab}iS_{Fij}^a (x-y),
\end{eqnarray}
where the flavor index $a,b=1,2,3$ or $u,d,s$, and $M_a=m+m_a$.
That is, the propagators of quarks are flavor-dependent.

The effective action describing meson interaction can be partially
obtained via loop effects of constituent quarks
\begin{eqnarray}\label{3.3}
e^{iS_{\rm eff}^q}&=&<0|T_qe^{i\int d^4x{\cal L}^{\rm
I}_\chi(x)}|0>
       \nonumber \\
 &=&\sum_{n=1}^\infty i\int d^4p_1\frac{d^4p_2}{(2\pi)^4}
  \cdots\frac{d^4p_n}{(2\pi)^4}\tilde{\Pi}_n(p_1,\cdots,p_n)
  \delta^4(p_1+p_2+\cdots+p_n) \nonumber \\
&\equiv&i\Pi_1(0)+\sum_{n=2}^\infty i\int \frac{d^4p_1}{(2\pi)^4}
  \cdots\frac{d^4p_{n-1}}{(2\pi)^4}\Pi_n(p_1,\cdots,p_{n-1}),
\end{eqnarray}
where $T_q$ is time-order product of constituent quark fields,
$\tilde{\Pi}_n(p_1,\cdots,p_n)$ is one-loop effects of constituent
quarks with $n$ external fields, $p_1,p_2,\cdots,p_n$ are their
four-momentum, and
\begin{equation}\label{3.4}
\Pi_n(p_1,\cdots,p_{n-1})=\int d^4p_n\tilde{\Pi}_n(p_1,\cdots,p_n)
  \delta^4(p_1+p_2+\cdots+p_n).
\end{equation}
Getting rid of all disconnected diagrams, we have
\begin{eqnarray}\label{3.5}
S_{\rm eff}&=&\sum_{n=1}^\infty S_n, \nonumber \\
S_1&=&\Pi_1^c(0),  \nonumber\\
S_2&=&\Pi_2^c(p)+\frac{F^2}{16}<\nabla_\mu U\nabla^\mu U^{\dag}>
   +\frac{1}{4}m_0^2<V_\mu V^{\mu}> \nonumber\\
S_n&=&\int \frac{d^4p_1}{(2\pi)^4}\cdots\frac{d^4p_{n-1}}
  {(2\pi)^4}\Pi_n^c(p_1,\cdots,p_{n-1}),
  \hspace{0.8in}(n\geq 3)
\end{eqnarray}
where "$c$" denotes "connected part", and two non-quark terms in
eq.(\ref{2.12}) have been added to obtain complete effective
action. Obviously, in eq.~(\ref{3.5}), the effective action
$S_{\rm eff}$ is expanded in powers of number of external vertex
and expressed as integral over external momentum. Hereafter we
shall call this method proper vertex expansion, and call $S_n$
$n$-point effective action. In terms of proper vertex expansion,
the effective actions include informations from all orders of
chiral expansion. That is, we can do resummation of momentum
expansion by this method. That is what we need.

For simplicity, we shall use the good approximation $m_u=m_d=0$,
which means the flavor index 1($u$) and 2($d$) are equivalent for
all kind of quantities. In ref. \cite{Wangvect} and
\cite{Wangrho}, what is mainly studied is the SU(2) sector of the
theory. So, $m_s$ plays no role in such kind of studies. Since
$m_u=m_d=0$, the effective actions can be put into simple forms
which are not flavor-related. For SU(3) case, we shall assume
$m_s\not=0$, which, as we can see later, is necessary for a
unitary theory when $\phi(1020)$ physics is considered. therefore,
the effective actions will be complicated and flavor-related.

The next two sections are calculations of some effective actions.
Sect.IV is about kinetic terms of vector mesons, and Sect.V is
about tree graphs for vector mesons decays.

\section{kinetic part of vector meson action}
\setcounter{equation}{0}

In the effective action, the kinetic part of vector mesons $S_{\rm
kin}$ can be derived from the two-point diagram as follow (fig.1).
Using Feynman rules generated from eq.(\ref{3.1}), We find that
the kinetic action is
\begin{figure}[hptb]
   \centerline{\psfig{figure=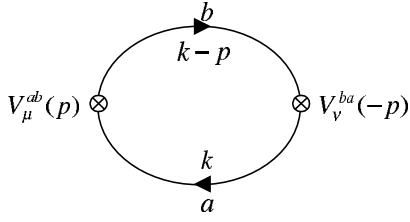,width=2.1in}}
 \centering
\begin{minipage}{5in}
   \caption{Two-point diagram of vector mesons.}
\end{minipage}
\end{figure}

\begin{eqnarray}\label{4.1}
S_{\rm kin}&=&\frac{i}{2}N_c\sum\limits_{ab}\int \frac{d^4 p_1 d^4
p_2}{(2\pi)^4}\delta(p_1+p_2)V_\mu^{ab}(p_1)
V_\nu^{ba}(p_2)T_{ab}^{\mu\nu}(p_2)
\nonumber\\
&=&\frac{1}{4}\sum\limits_{ab}\int
\frac{d^4p}{(2\pi)^4}V_\mu^{ab}(p)V_\nu^{ba}(-p)
[a_1g^{\mu\nu}-a_2g^{\mu\nu}p^2+b_2p^\mu
p^\nu+a_3(p^2)g^{\mu\nu}p^2- b_3(p^2)p^\mu p^\nu].
\end{eqnarray}
Here $V_\mu^{ab}$ refers to the $(ab)$ element of matrix $V_\mu$
~(\ref{2.11}),
and
\begin{eqnarray}\label{4.2}
T_{ab}^{\mu\nu}(p)=\mu^{\ep}\int \frac{d^d
k}{(2\pi)^d}\frac{Tr[\gamma^\mu(\sla{k}-\sla{p}+M_b)\gamma^\nu
(\sla{k}+M_a)]}{[(k-p)^2-M_b^2](k^2-M_a^2)}.
\end{eqnarray}
The coefficients in the second line of eq.~(\ref{4.1}) are
\begin{eqnarray}\label{4.3}
a_1  &=&\frac{3}{2}g^2(M_a-M_b)^2-\frac{N_c}{2\pi^2}\int_0^1 dx
  \ln\frac{M_a^2(1-x)+M_b^2x}{m^2}\left[M_a^2(1-x)+M_b^2x-M_aM_b\right],
\nonumber \\
a_2  &=&g^2-\frac{N_c}{2\pi^2}\int_0^1 dx
   x(1-x)\left[1-\frac{M_aM_b}{M_a^2(1-x)+M_b^2x}+
   2\ln\frac{M_a^2(1-x)+M_b^2x}{m^2}\right], \nonumber \\
b_2  &=&g^2-\frac{N_c}{\pi^2}\int_0^1 dx
    x(1-x)\ln\frac{M_a^2(1-x)+M_b^2x}{m^2},\nonumber \\
a_3(p^2)&=&\frac{N_c}{\pi^2}\int_0^1
        dx\left[x(1-x)\ln\left(1-\frac{p^2x(1-x)}{M_a^2(1-x)+M_b^2x}
    \right)-
        \frac{R(p^2)}{2p^2}[M_a^2(1-x)+M_b^2x-M_aM_b]\right],
    \nonumber \\
b_3(p^2)&=&\frac{N_c}{\pi^2}\int_0^1
         dx
x(1-x)\ln\left(1-\frac{p^2x(1-x)}{M_a^2(1-x)+M_b^2x}\right),
\nonumber \\
R(p^2)&=&\ln\left(1-\frac{p^2x(1-x)}{M_a^2(1-x)+M_b^2x}\right)+
       \frac{p^2x(1-x)}{M_a^2(1-x)+M_b^2x}.
\end{eqnarray}
Because $R(p^2)$ is of $O(p^4)$, we find that $a_3(p^2)\sim
O(p^2)$, $b_3(p^2)\sim O(p^2)$. The constant $g$ in
eq.~(\ref{4.3}) is a universal coupling constant, which absorbs
the logarithmic divergence originating from quark loop integral,
\begin{eqnarray}\label{4.4}
\frac{3}{8}g^2=\frac{N_c}{(4\pi)^{d/2}}\left(\frac{\mu^2}{m^2}
\right)^{\ep/2}
  \Gamma\left(2-\frac{d}{2}\right).
\end{eqnarray}

At chiral limit ($m_u=m_d=m_s=0$), we can find that
\begin{eqnarray}\label{4.5}
a_1  &=&0,\nonumber \\
a_2  &=&b_2  =g^2,\nonumber \\
a_3(p^2)&=&b_3(p^2)=\frac{N_c}{\pi^2}\int_0^1
         dx x(1-x)\ln\left(1-\frac{p^2x(1-x)}{m^2}\right).
\end{eqnarray}
Therefore, $S_{\rm kin}$ is close to the standard form
\begin{eqnarray}\label{4.6}
S_{\rm kin}^{\rm stand}=\frac{1}{4}\sum\limits_{ab}\int
\frac{d^4p}{(2\pi)^4}V_\mu^{ab}(p)V_\nu^{ba}(-p)
[-g^{\mu\nu}p^2+p^\mu p^\nu+m_{_Vab}^2g^{\mu\nu}],
\end{eqnarray}
provided we rescale $V_\mu\rightarrow V_\mu/\sqrt{a_2}=V_\mu/g$. While beyond
the chiral limit, $a_2\not=b_2$, so the $-g^{\mu\nu}p^2$
 term and the $p^\mu p^\nu$ term have no common coefficients. Gauge
symmetry is thus broken. But we don't need to bother about this,
because throughout this paper, the condition $\pa_\mu V^\mu=0$ is
used. The $p^\mu p^\nu$ term can thus be discarded.

Another problem is the higher derivatives in eq.~(\ref{4.1}),
which makes the vector mesons ill-defined. Fortunately, because
the vector meson field is external, we can make a
momentum-dependent transformation $V_\mu (p)\rightarrow
(1+f(p^2)p^2)V_\mu (p)$ such that, in the final form, terms with
derivatives higher than 2 vanish, i.e.
\begin{eqnarray}\label{4.7}
a_3(p^2)-2f_1a_1+(a_1-a_2p^2+a_3(p^2)p^2)(2f(p^2)+f^2(p^2)p^2)=0,
\end{eqnarray}
where $f_1=f(p^2)|_{p^2=0}$. If $a_1\not=0$ (i.e. $m_a\not=m_b$),
eq.~(\ref{4.7}) can't determine the value of $f_1$. However, when
$m_a=m_b$, we have $a_1=0$, and $f_1$ can be determined as
$f_1=a^1_3/(2a_2)|_{m_a=m_b}$, where $a^1_3=a_3(p^2)|_{p^2=0}$.
In the case of $m_a\not=m_b$, because $a_1$
is very small, $f_1$ changes slightly. $a^1_3/(2a_2)|_{m_a=m_b}$
is thus a good approximation for $f_1$ at this case. Substituting
$(m_a+m_b)/2$ for the argument $m_a$ in this expression, we obtain
\begin{eqnarray}\label{4.8}
f_1=-N_c\left\{15\pi^2(M_a+M_b)^2\Big[g^2-\frac{N_c}{6\pi^2}
\ln\frac{(M_a+M_b)^2}{4m^2}\Big]\right\}^{-1}.
\end{eqnarray}
Then we get the transformation factor
\begin{eqnarray}\label{4.9}
\alpha(p^2)\equiv 1+f(p^2)p^2=\sqrt{1+\frac{(2f_1a_1-a_3(p^2))p^2}
{a_1-a_2p^2+a_3(p^2)p^2}}.
\end{eqnarray}

After this transformation, we find that
\begin{eqnarray}\label{4.10}
S_{\rm kin}=\frac{1}{4}\sum\limits_{ab}\int
\frac{d^4p}{(2\pi)^4}V_\mu^{ab}(p)V^{\mu ba}(-p)
(a_1-a_2p^2+2f_1a_1p^2).
\end{eqnarray}
Comparing it with the standard form, we find the proper rescaling
\begin{eqnarray}\label{4.11}
V_\mu\rightarrow \frac{V_\mu}{\td g} \equiv \frac{V_\mu}{\sqrt{a_2-2f_1a_1}}.
\end{eqnarray}
The physical pseudoscalar meson fields can be
obtained via field rescaling $\Phi\rightarrow 2F_\Phi^{-1}\Phi$
($\Phi=\pi,K,\eta_8$).

What should be noted is that, the rescaling factor $1/\td g$ and
the transformation factor $\alpha(p^2)$ are both flavor-related.
They are different for different vector mesons (for $\rho$, the
rescaling factor is $1/\td g=1/g$). In the following expressions,
We shall omit factors as $1/\td g$, $\alpha(p^2)$ and $2/F_\Phi$
for external lines, and include them only in final results.

\section{Quark Loop Contributions to vector meson decays}
\setcounter{equation}{0}

\subsection{Effective actions}

For vector mesons decays, we should include the two-point and
three-point diagram for tree level of effective actions. The
space-like condition of vector meson $\pa^\mu V_\mu=0$ is used to
simplify the calculations.

For two-point diagram, we should calculate fig.2(a). The
corresponding effective action reads
\begin{figure}[hptb]
   \centerline{\psfig{figure=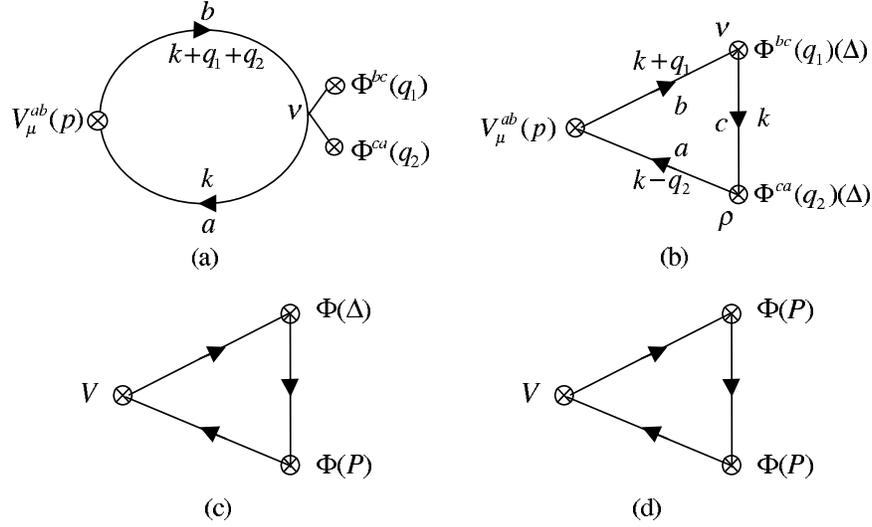,width=4.5in}}
 \centering
\begin{minipage}{5in}
   \caption{Two-point and three-point diagrams for effective actions.}
\end{minipage}
\end{figure}

\begin{eqnarray}\label{5.1}
S_2&=&\frac{iN_c}{8}\sum\limits_{abc} \int \frac{d^4p d^4q_1
d^4q_2}{(2\pi2\pi)^4}\delta(p+q_1+q_2)
V_\mu^{ab}(p)\Phi^{bc}(q_1)\Phi^{ca}(q_2)(q_2-q_1)_\nu
T_{ab}^{\mu\nu}(q_1+q_2)\nonumber \\
&=&\sum\limits_{abc}\int \frac{d^4p d^4q_1 d^4q_2}{(2\pi2\pi)^4}
\delta(p+q_1+q_2) V_\mu^{ab}(p)\Phi^{bc}(q_1)\Phi^{ca}(q_2)q_2^\mu
A_{ab}(p^2),
\end{eqnarray}
where
\begin{eqnarray}\label{5.2}
A_{ab}(p^2)&=&\frac{3}{16}g^2(M_a-M_b)^2-\frac{1}{8}g^2 p^2
-\frac{N_c}{16\pi^2}\int_0^1 dx
   \ln\frac{D_1^{ab}}{m^2}[M_a^2(1-x)+M_b^2x-M_aM_b-2x(1-x)p^2],
   \nonumber\\
D_1^{ab}(p^2)&=&M_a^2(1-x)+M_b^2x-p^2x(1-x).
\end{eqnarray}
Contributions from $S'$ vanishes here.

For three-point diagram, we should include three kinds of vertices:
$V-\Delta\Delta$,$V-\Delta P$,$V-PP$. The contribution from the first
vertex (fig.2(b)) is
\begin{eqnarray}\label{5.3}
S_3^{\Delta}&=&\frac{i}{4}N_c g_A^2\sum\limits_{abc}\int
\frac{d^4p d^4q_1 d^4q_2}{(2\pi2\pi)^4}\delta(p+q_1+q_2)
V_\mu^{ab}(p)\Phi^{bc}(q_1)\Phi^{ca}(q_2)q_{1\nu}q_{2\rho}
T_{ab-c}^{\mu\nu\rho}(q_1,q_2)\nonumber\\
&=&\sum\limits_{abc}\int\frac{d^4p d^4q_1 d^4q_2}{(2\pi2\pi)^4}
\delta(p+q_1+q_2) V_\mu^{ab}(p)\Phi^{bc}(q_1)\Phi^{ca}(q_2)q_2^\mu
g_{A}^2B_{ab-c}(p^2,q_1^2,q_2^2),
\end{eqnarray}
Here, we have used the definitions that
\begin{eqnarray}\label{5.4}
T_{abc}^{\mu\nu\rho}(q_1,q_2)&=&\mu^{\ep}\int
\frac{d^d k}{(2\pi)^d}\frac{Tr[\gamma^\mu(\sla{k}+\sla{q}_1+M_b)
\gamma^\nu
(\sla{k}+M_c)\gamma^\rho(\sla{k}-\sla{q}_2+M_a)]}
{[(k+q_1)^2-M_b^2](k^2-M_c^2)[(k-q_2)^2-M_a^2]},\nonumber\\
T_{ab-c}^{\mu\nu\rho}&=&T_{abc}^{\mu\nu\rho}(M_c\rightarrow
-M_c).
\end{eqnarray}
and
\begin{eqnarray}\label{5.5}
B_{abc}(p^2,q_1^2,q_2^2)&=&-\frac{1}{8}\left[\left(\frac{N_c}
{6{\pi}^2}
-g^2\right)p^2
+\frac{N_c}{2{\pi}^2}\int_0^1xdx\int_0^1dy
\left(\alpha_1\ln\frac{D_2^{abc}}{m^2}+\frac{\alpha_2^{abc}}
{D_2^{abc}}\right)
\right],\nonumber\\
\alpha_1&=&p^2(1-xy)-q_1^2(1-x-xy)-q_2^2x(1-2y),\nonumber\\
\alpha_2^{abc}&=&-p^2M_c[M_ax(1-y)+M_b(1-x)]+q_1^2M_a[M_b(1-x)-M_cxy]+
   q_2^2M_bx[M_a(1-y)-M_cy]\nonumber\\
&&+p^2q_1^2x(1-x)^2+p^2q_2^2x^2(1-x+xy)(1-y)^2+q_1^2q_2^2x^2y^2(1-xy),
\nonumber\\
D_2^{abc}&=&M_a^2x(1-y)+M_b^2(1-x)+M_c^2xy-p^2x(1-x)(1-y)-q_1^2xy(1-x)
-q_2^2x^2y(1-y),\nonumber\\
B_{ab-c}&=&B_{abc}(M_c\rightarrow -M_c).
\end{eqnarray}

The contributions from the latter two vertices (fig.2(c) and (d)) are
\begin{eqnarray}\label{5.6}
S_3^P=\sum\limits_{abc}\int\frac{d^4p d^4q_1 d^4q_2}{(2\pi2\pi)^4}
\delta(p+q_1+q_2) V_\mu^{ab}(p)\Phi^{bc}(q_1)\Phi^{ca}(q_2)q_2^\mu
B_{abc}'(p^2,q_1^2,q_2^2),
\end{eqnarray}
where $B'$ is defined as
\begin{eqnarray}\label{5.7}
V_\mu^{ab}(p)&&\Phi^{bc}(q_1)\Phi^{ca}(q_2)B_{abc}'(p^2,q_1^2,q_2^2)=
\frac{\kappa m_s N_c}{(4\pi)^2}\int_0^1 xdx\int_0^1dy\nonumber\\
&&\Big\{V_\mu^{ab}(p)\Phi^{bc}(q_1)\td{\Phi}^{ca}(q_2)g_{A}
\Big[M_c-M_a+M_b+
\Big(\frac{6\pi^2g^2}
{N_c}-\ln\frac{D_2^{abc}}{m^2}\Big)(M_a-2M_b-M_c)
-\frac{\beta_1^{abc}}
{D_2^{abc}}\Big]\nonumber\\
&&+V_\mu^{ab}(p)\td{\Phi}^{bc}(q_1)\Phi^{ca}(q_2)g_{A}
\Big[M_a-M_b+M_c+
\Big(\frac{6\pi^2g^2}
{N_c}-\ln\frac{D_2^{abc}}{m^2}\Big)(M_b-2M_a-M_c)-\frac{\beta_2^{abc}}
{D_2^{abc}}\Big]\nonumber\\
&&+V_\mu^{ab}(p)\td{\Phi}^{bc}(q_1)\td{\Phi}^{ca}(q_2) \kappa
m_s\Big[1+xy-(1+3xy)\Big(\frac{6\pi^2g^2}
{N_c}-\ln\frac{D_2^{abc}}{m^2}\Big)-\frac{\beta_3^{abc}}
{D_2^{abc}}\Big]\Big\},
\end{eqnarray}
with
\begin{eqnarray}\label{5.8}
\beta_1^{abc}&=&M_aM_bM_c+p^2x(1-y)[M_ax(1-y)+M_b(1-x)+M_c(1-x+xy)]
\nonumber\\
&&+q_1^2xy[M_a(1-xy)+M_b(1-x)+M_cxy]+q_2^2M_bx^2y(1-y)\nonumber\\
\beta_2^{abc}&=&M_aM_bM_c+p^2(1-x)[M_ax(1-y)+M_b(1-x)+M_cx]\nonumber\\
&&+q_1^2M_axy(1-x)+q_2^2xy[M_ax(1-y)+M_b(1-xy)+M_cxy]\nonumber\\
\beta_3^{abc}&=&M_aM_b(1-xy)+(M_a+M_b)M_cxy+p^2x(1-x)(1-y)(1+xy)+
q_1^2x^2y^2(1-x)+q_2^2x^3y^2(1-y).
\end{eqnarray}
Here,
\begin{eqnarray}\label{5.9}
\td{\Phi}=\left(\begin{array}{ccc}
           0 &0 &\Phi^{13}\\
           0 &0 &\Phi^{23}\\
           \Phi^{31} &\Phi^{32} &2\Phi^{33}
          \end{array} \right)=
\sqrt{2}
\left(\begin{array}{ccc}
           0 &0 &K^+   \\
         0 &0  &K^0   \\
       K^-&\bar{K}^0&-\frac{4}{\sqrt{6}}\eta_8
       \end{array} \right),
\end{eqnarray}
which results from the simplified ${\cal M}={\rm
diag}\{0,0,m_s\}$. Note that the form of $B'$ is not given
directly, which depends on the flavor index. For $\rho$ and
$\omega$ decay, $B'=0$.

Thus the total action from quark loop is
\begin{eqnarray}\label{5.10}
S_{\rm quark}= \sum\limits_{abc}\int\frac{d^4p d^4q_1
d^4q_2}{(2\pi2\pi)^4}\delta(p+q_1+q_2)
V_\mu^{ab}(p)\Phi^{bc}(q_1)\Phi^{ca}(q_2)q_2^\mu
f^{(0)}_{abc}(p^2,q_1^2,q_2^2),
\end{eqnarray}
where
\begin{eqnarray}\label{5.11}
f^{(0)}_{abc}(p^2,q_1^2,q_2^2)=A_{ab}(p^2)+g_{A}^2B_{ab-c}(p^2,q_1^2,q_2^2)+
     B_{abc}'(p^2,q_1^2,q_2^2).
\end{eqnarray}
Now we have resumed all orders of momentum expansion, which is
embodied by $D_1^{ab}$ and $D_2^{abc}$. To obtain the leading
order of momentum expansion, we should first take chiral limit
$m_u=m_d=m_s=0$. But we should remember that, as indicated at the
end of Sec.III, it is inconsistent to study $\phi$ physics at
chiral limit.

\subsection{Vector meson dominant and KSRF sum rules}

We also care about decay $\rho\rightarrow e^+ e^-$. The direct
coupling between photon and vector meson resonances is also
yielded by the effects of quark loops. In chiral limit, the VMD
vertex at the leading order of large $N_c$ expansion, after the
condition $\pa^\mu V_\mu=0$ is applied, reads
\begin{eqnarray}\label{5.12}
 {S}_{\rm VMD}=-\frac{e}{2}\int\frac{d^4p}{(2\pi)^4}
 A_\mu(-p)<QV^\mu (p)>f_{\rho\gamma}^{(0)}(p^2)p^2,
\end{eqnarray}
where $A^\mu$ is photon field, $Q={\rm diag}\{2/3,-1/3,-1/3\}$
is charge operator of quark fields, and
\begin{eqnarray}\label{5.13}
f_{\rho\gamma}^{(0)}(p^2)=-\frac{8A_{11}(p^2)}{gp^2}=
g-\frac{N_c}{g\pi^2}\int_0^1dx\cdot x(1-x)
  \ln(1-\frac{x(1-x)p^2}{m^2}).
\end{eqnarray}

In addition, at the leading order of large $N_c$ expansion, the
$V-\Phi\Phi$ vertex (where $V$ stands for vector mesons and $\Phi$ for
pseudoscalar mesons) in chiral limit
reads from eq.~(\ref{5.10})
\begin{eqnarray}\label{5.14}
S_{\rm V\Phi\Phi}= -\frac{1}{2}\int\frac{d^4p d^4q_1
d^4q_2}{(2\pi2\pi)^4}\delta(p+q_1+q_2)
<V_\mu(p)\Phi(q_1)\Phi(q_2)>q_2^\mu f_{\rho\pi\pi}^{(0)}(p^2)p^2,
\end{eqnarray}
where
\begin{eqnarray}\label{5.15}
f_{\rho\pi\pi}^{(0)}(p^2)=
-\frac{8f^{(0)}_{111}(p^2,0,0)}{gF_\pi^2p^2}.
\end{eqnarray}
The rescaling factors $1/g$ for vector mesons and $2/F_\pi$ for
pseudoscalar mesons is included both in eq.~(\ref{5.13})
and eq.~(\ref{5.15}).

It is well known that the KSRF(I) sum rule \cite{KSRF}
\begin{equation}\label{5.16}
 f_{\rho\gamma}(m_\rho^2)=f_{\rho\pi\pi}(m_\rho^2)F_\pi^2
\end{equation}
is the result of current algebra and PCAC. We
expect it to be valid at the leading order of large $N_c$ expansion.
Therefore, the KSRF(I) sum rule is satisfied when
$B_{11-1}(p^2,0,0)\big|_{p^2=m_\rho^2}=0$, or, using definition
~(\ref{5.5}),
\begin{eqnarray}\label{5.17}
g^2=\frac{N_c}{2\pi^2}\int_0^1dx\int_0^1dy\cdot
      x(1-xy)\Big[1+\frac{m^2}{m^2-x(1-x)(1-y)p^2}+
      \ln\frac{m^2-x(1-x)(1-y)p^2}{m^2}\Big]\Big|_{p^2=m_\rho^2}.
\end{eqnarray}
Setting $m\simeq 460$MeV (see Appendix A), we find that $g=0.329$.

\section{Goldstone boson loop contributions to vector meson decays}
\setcounter{equation}{0}

In this section, we shall calculate one-loop correction of
pseudoscalar meson to vector mesons decays and provide
a complete prediction on these reactions.
Because the contribution of loop effects is suppressed by $N_c^{-1}$
expansion, we shall use some approximations for simplicity.

Firstly, because the quark masses in meson loop are doubly
suppressed, we shall discard $S'$ and $\kappa P$ parts in ${\cal
L}_{\chi}^I$, because they are of $O(m_s)$. This means $B'=0$ when
tree-level action is used in meson-loop calculations. Secondly, in
propagators of pseudoscalar mesons, since $m_\pi^2\ll m_{K}^2$, we
assume that pion is a massless particle.

There is another approximation we have used. In dimensional regularization,
$\int d^{d}kk^{2n}\equiv 0\;(n\geq -1)$, Thus, we have
$$\int d^{d}k\frac{k^{2s}}{k^2-m_\Phi^2+i\epsilon}\propto
m_\Phi^{2s} \times{\rm quadratically\;divergent\;term},\hspace{1in}
(s\geq1).$$
Therefore, in this paper we shall ignore all contributions from quartic
divergences or higher order ones. As we shall see, in the following
calculation the lowest order divergence is quadratic. It means that we
can take the approximation $q^2=0$ ($q^\mu$ is the four-momentum
of pseudoscalar mesons) in calculation on pseudoscalar meson loops.

At the order of $N_c^{-1}$ expansion next to the leading one,
there are three kinds of loop diagrams of pseudoscalar mesons need
to be calculated (fig.3). As to the tadpole diagram in fig.3(a)
and (d), we only include contributions from $K$ and $\eta_8$
mesons, since the massless $\pi$ meson does not contribute due to
the relation $\int d^{d}kk^{2n}\equiv 0\;(n\geq -1)$. In addition,
when we calculate two-point diagram (fig.3(b)(e)), we must include
all chain-like diagrams of pseudoscalar meson loops which have
imaginary part (fig.3(c)(f)), e.g., $\pi\pi$ loop for $\rho$
decay. Because it will generates a large imaginary part of
$S$-matrix for vector mesons physics.

\begin{figure}[hptb]
   \centerline{\psfig{figure=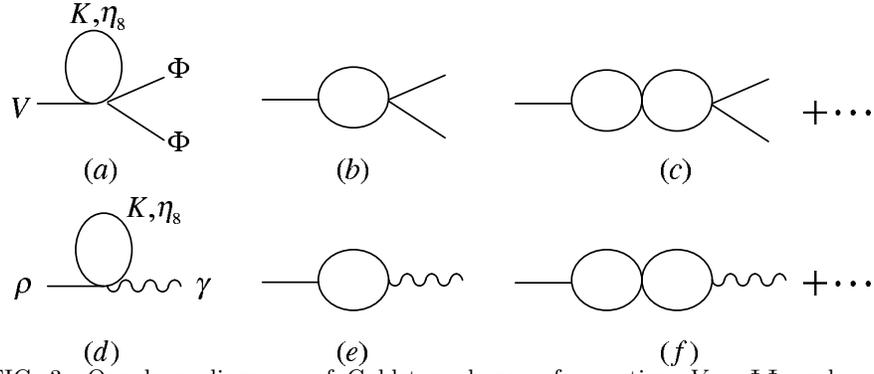,width=4.5in}}
 \centering
\begin{minipage}{5in}
   \caption{One loop diagrams of Goldstone bosons for vertices $V-\Phi\Phi$ and
   $\rho-\gamma$. (a)(d)Tadpole diagram of $K$ and $\eta_8$.
   (b)(e)Two-point diagram of pseudoscalar mesons.
   (c)(f)Chain-like diagrams of pseudoscalar meson loops.}
\end{minipage}
\end{figure}

\subsection{Tadpole Diagram}

When calculating tadpole, we need various quark one-loop diagrams with
various external vertices but with one vector meson external source and
four pseudoscalar meson external sources. They are shown in fig.4.

As we can see later, the contributions from fig.(e) and (f) can be
omitted, so we have
\begin{eqnarray}\label{6.1}
S_{V-4\Phi}=\sum\limits_{abcde}\int\frac{d^4p d^4q_1\cdots d^4q_4}
{(2\pi)^{4\times 4}}\delta(p+q_1+\cdots+q_4)
V_\mu^{ab}(p)\Phi^{bc}(q_1)\Phi^{cd}(q_2)\Phi^{de}(q_3)\Phi^{ea}(q_4)
f^{\mu}_{abcde}(q_1,\cdots,q_4),
\end{eqnarray}
where $f^{\mu}$ includes contributions from fig. (a)$\sim$(d):
\begin{eqnarray}\label{6.2}
f^{\mu}_{abcde}(q_1,\cdots,q_4)&=&\frac{i}{2^7\times 3}N_c
     (q_1-3q_2+3q_3-q_4)_\nu T_{ab}^{\mu\nu}(q_1+\cdots+q_4)
     \nonumber \\
&&-\frac{i}{64}N_c(q_2-q_1)_\nu(q_4-q_3)_\rho
      T_{abd}^{\mu\nu\rho}(q_1+q_2,q_3+q_4) \nonumber \\
&&-\frac{i}{96}N_c g_{A}^2 [q_{1\nu} (q_2-2q_3+q_4)_{\rho}
      T_{ab-c}^{\mu\nu\rho}(q_1,q_2+q_3+q_4) \nonumber \\
       &&\;\;\;\;\;\;\;\;\;\;\;\;\;\;\;\;\;
       +(q_1-2q_2+q_3)_\nu q_{4\rho}
      T_{ab-e}^{\mu\nu\rho}(q_1+q_2+q_3,q_4)].
\end{eqnarray}
Here, we have omitted  $S'$ and $\kappa P$ in ${\cal L}_{\chi}^I$.

Now, arbitrary two of the four $\Phi$ should be contracted as long
as they are ${\eta}_8$ or $K$. Using the propagators for them in
momentum space such as
\begin{figure}[hptb]
   \centerline{\psfig{figure=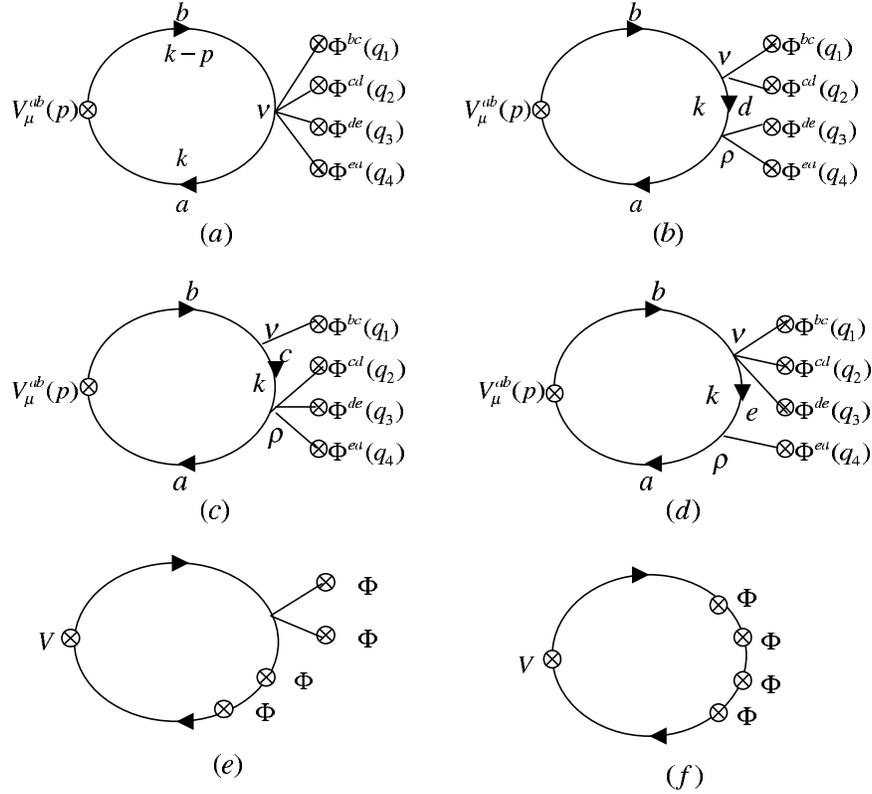,width=4.5in}}
 \centering
\begin{minipage}{5in}
   \caption{Diagrams of quark loop for vertex $V-4\Phi$.}
\end{minipage}
\end{figure}
\begin{eqnarray}\label{6.3}
\stackrel{\sct{0.5}}{K^+(q_1)K^-}(q_2)=(2\pi)^4\delta(q_1+q_2)
\frac{i}{q_1^2-m_{K}^2+i\ep},
\end{eqnarray}
(remembering $m_{K}=m_{\eta_8}$) and considering all possible
contractions, we have
\begin{eqnarray}\label{6.4}
S_{\rm tad}&=&2\int\frac{d^4p d^4q_1 d^4q_2}{(2\pi
2\pi)^4}\delta(p+q_1+q_2)
V_\mu^{13}(p)\Phi^{32}(q_1)\Phi^{21}(q_2)
\int\frac{d^4q}{(2\pi)^4}\frac{i}{q^2-m_{K}^2}\nonumber\\
&&\Big [\frac{1}{6}f^{\mu}_{13211}(q_1,q_2,q,-q)
+f^{\mu}_{13213}(q_1,q_2,q,-q)
+\frac{1}{6}f^{\mu}_{13221}(q_1,q,q_2,-q)
+\frac{1}{6}f^{\mu}_{13222}(q_1,q,q_2,-q)
\nonumber\\
&&+f^{\mu}_{13232}(q_1,q,-q,q_2)-\frac{2}{6}
f^{\mu}_{13321}(q,q_1,q_2,-q)
-\frac{2}{6}f^{\mu}_{13322}(q,q_1,-q,q_2)
+2f^{\mu}_{13132}(q,-q,q_1,q_2)
\nonumber\\
&&+\left(\frac{2}{\sqrt{6}}\right)^2f^{\mu}_{13332}(q,-q,q_1,q_2)\Big ]
\end{eqnarray}
for the vertex $K^{*+}(p)\bar{K}^0(q_1)\pi^-(q_2)$ and similar
expressions for the vertices $\rho(p)\pi(q_1)\pi(q_2)$ and
$\phi(p)K(q_1)K(q_2)$.
When expanding
the expression in the brackets in powers of $q^\mu$, and
discarding the odd-order terms, we obtain a polynomial in powers
of $q^2$. Because the two contracted $\Phi$s are external fields
as far as the quark loop concerned, due to the approximation
$q^2=0$, what we do amounts to setting $q_\mu=0$ in the bracket.
(Now let we turn to the
contribution of fig.(e) to $f^\mu$. It must be of this form:
$(q_2-q_1)_\nu q_{3\rho} q_{4\sigma}
T_{abcd}^{\mu\nu\rho\sigma}(q_1+q_2,q_3,q_4)$. After contraction,
two of the four $q$s should be set to be $0$. This term is thus
vanished. Similar argument is valid for fig.(f).) The calculations
are thus simplified. Eventually, we get
\begin{eqnarray}\label{6.5}
S_{\rm tad}&=&\int\frac{d^4p d^4q}{(2\pi 2\pi)^4}
V_\mu^{ab}(p)\Phi^{bc}(-p-q)\Phi^{ca}(q)q^\mu C_{abc}(p^2)
\end{eqnarray}
with
\begin{eqnarray}\label{6.6}
C_{122}(p^2)&=&\frac{\lambda m_{K}^2}{6(4\pi)^2 F_{K}^2}
[3B_{113}(p^2)-3g_{A}^2 B_{11-1}(p^2)-5f^{(0)}_{111}(p^2)]\nonumber\\
C_{132}(p^2)&=&\frac{\lambda m_{K}^2}{6(4\pi)^2 F_{K}^2}
[3B_{133}(p^2)-3g_{A}^2 B_{13-1}(p^2)-10f^{(0)}_{131}(p^2)]\nonumber\\
C_{33c}(p^2)&=&\frac{3\lambda m_{K}^2}{4(4\pi)^2 F_{K}^2}
[B_{331}(p^2)-g_{A}^2
B_{33-1}(p^2)-3f^{(0)}_{331}(p^2)]\;\;\;\;\;\;(c=1,2)
\end{eqnarray}
for vertex $\rho^+\pi^0\pi^-$, $K^{*+}\bar{K}^0\pi^-$ and $\phi K
K$ respectively, where $B(p^2)=B(p^2,0,0)$, and
\begin{eqnarray}\label{6.7}
f^{(0)}_{abc}(p^2)=A_{ab}(p^2)+g_{A}^2 B_{ab-c}(p^2)
\end{eqnarray}
is the simplified form of $f^{(0)}_{abc}(p^2,q_1^2,q_2^2)$ according to
$q^2=0$. Here, we have neglected the differences between
$m_{K}$ and $m_{\eta}$, and between $F_{K}$
and $F_{\eta}$, since they are doubly suppressed by light quark mass expansion
and $N_c^{-1}$ expansion. The constant $\lambda$ absorbs the divergence:
\begin{eqnarray}\label{6.8}
\lambda=\left(\frac{4\pi\mu^2}{m_{K}^2}\right)^{\ep/2}
\Gamma\left(1-\frac{d}{2}\right).
\end{eqnarray}

After similar consideration, we can find the tadpole loop
corrections of $K$ or $\eta_8$ mesons to VMD vertex (see fig.3(d)) is
\begin{eqnarray}\label{6.9}
{S}_{\rm VMD}^{(t)}=\frac{e\lambda m_{K}^2}
  {16\pi^2F_{K}^2}\int\frac{d^4p}{(2\pi)^4}
A_\mu(-p)\rho^{0\mu}(p)f_{\rho\gamma}^{(0)}(p^2)p^2.
\end{eqnarray}

In Appendix B, $\lambda=0.54$ is determined.

\subsection{Two-Point Diagram and Chain-Like Approximation}

For two-point diagram, we need the vertex of $4\Phi$, which in
principle should be generated from quark loop. For simplicity, we
can alternatively obtain it from the effective Lagrangian of ChPT at
order $p^2$ and order $p^4$,
\begin{eqnarray}\label{6.10}
{\cal
L}_{4\Phi}=&&\frac{F_0^2}{96}(<\pa_\mu\Phi\Phi\pa^\mu\Phi\Phi>-
<\pa_\mu\Phi\pa^\mu\Phi\Phi^2>)\nonumber\\
&&+L_1<\pa_\mu\Phi\pa^\mu\Phi><\pa_\nu\Phi\pa^\nu\Phi>
+L_2<\pa_\mu\Phi\pa_\nu\Phi><\pa^\mu\Phi\pa^\nu\Phi>
+L_3<\pa_\mu\Phi\pa^\mu\Phi\pa_\nu\Phi\pa^\nu\Phi>,
\end{eqnarray}
with the coefficients determined by ChQM (see Appendix A)
\begin{eqnarray}\label{6.11}
L_1=\frac{\gamma}{24},\;\;\;\;\;\;\;L_2=\frac{\gamma}{12},\;\;\;\;\;\;\;
L_3=\frac{\gamma}{12}(g_{A}^4-3),\;\;\;\;\;\;\;
\gamma=\frac{N_c}{(4\pi)^2}.
\end{eqnarray}
At chiral limit, it is known that the renormalized
$F_0$ is just the decay constant of $\pi$ mesons: $F_0=F_\pi$.

As to the vertex $V-\Phi\Phi$, it is the sum of
two- and three-point diagrams of quark loops:
\begin{eqnarray}\label{6.12}
{\cal L}_{V-2\Phi}=i\sum\limits_{abc}f^{(0)}_{abc}(-{\na}^2)V_\mu^{ab}(x)
\Phi^{bc}(x)\pa ^{\mu}\Phi^{ca}(x),
\end{eqnarray}
where ${\na}^2$ acts only on the coordinates of $V_\mu$.
Thus, the effective
Lagrangian is
${\cal L}_{V-2\Phi}+{\cal L}_{4\Phi}$.
Contracting two $\Phi$s in ${\cal L}_{V-2\Phi}$ with any two $\Phi$s in
${\cal L}_{4\Phi}$, and summing over all possible contractions of $K$,
$\eta_8$ and $\pi$ (see fig.3(b)), we have
\begin{eqnarray}\label{6.13}
S_{\rm TPD}=\sum\limits_{abc}\int \frac{d^4p d^4q}{(2\pi2\pi)^4}
V_\mu^{ab}(p)\Phi^{bc}(-p-q)\Phi^{ca}(q)q^{\mu}E_{abc}(p^2),
\end{eqnarray}
with
\begin{eqnarray}\label{6.14}
E_{122}(p^2)&=&f^{(0)}_{111}(p^2)\Sigma_\pi(p^2)+f^{(0)}_{113}(p^2)\Sigma_K(p^2)\nonumber \\
E_{132}(p^2)&=&f^{(0)}_{131}(p^2)\Sigma_{K\pi}(p^2)+
\frac{1}{2}\Big(f^{(0)}_{131}(p^2)+2f^{(0)}_{133}(p^2)\Big)\Sigma_K(p^2)
\nonumber \\
E_{33c}(p^2)&=&3f^{(0)}_{331}(p^2)\Sigma_K(p^2)\nonumber \\
\end{eqnarray}
for vertex $\rho^+\pi^0\pi^-$, $K^{*+}\bar{K}^0\pi^-$ and $\phi K
K$ respectively, where
\begin{eqnarray}\label{6.15}
\Sigma_\pi(p^2)&=&-\frac{F_0^2p^2}{4\pi^2F_{\pi}^4}\Big(1+\frac{3p^2}{4F_0^2}
  (g^2-\frac{g_A^4}{3\pi^2})\Big)\Big[ \frac{\lambda}{6} +\int_{0}^{1}dx
x(1-x)\ln\frac{
x(1-x)p^2}{m_{K}^2}\nonumber\\
&&\;\;\;\;\;\;\;\;\;\;\;\;\;\;\;\;\;\;\;\;\;\;\;\;\;\;\;\;\;\;\;\;\;\;\;\;\;
\;\;\;\;\;\;\;\;\;\;\;\;\;
+\frac{i}{6}{\rm Arg}(-1)\theta(p^2-4
m_\pi^2)\Big( 1-\frac{4m_\pi^2}{p^2}
\Big)^{3/2}\Big]\nonumber \\
\Sigma_K(p^2)&=&\frac{F_0^2m_K^2}{8\pi^2F_K^4}\Big(1+\frac{3p^2}{4F_0^2}
  (g^2-\frac{g_A^4}{3\pi^2})\Big)\Big[
\lambda(1-\frac{p^2}{6m_K^2})+\int_{0}^{1}dx\Big(1-x(1-x)\frac{p^2}{m_K^2}\Big)
\ln\Big(1-x(1-x)\frac{p^2}{m_{K}^2}\Big)\Big]
\nonumber\\
&=&\frac{F_0^2m_K^2}{8\pi^2F_K^4}\Big(1+\frac{3p^2}{4F_0^2}
  (g^2-\frac{g_A^4}{3\pi^2})\Big)\Big[
\lambda(1-\frac{p^2}{6m_K^2})+\int_{0}^{1}dx\Big(1-x(1-x)\frac{p^2}{m_K^2}\Big)
\ln\left|1-x(1-x)\frac{p^2}{m_{K}^2}\right|\nonumber\\
&&\;\;\;\;\;\;\;\;\;\;\;\;\;\;\;\;\;\;\;\;\;\;\;\;\;\;\;\;\;\;\;\;
\;\;\;\;\;\;\;\;\;\;\;\;\;\;\;\; -\frac{i}{6}{\rm
Arg}(-1)\theta(p^2-4m_{K}^2)\frac{p^2}{m_K^2}\Big(
1-\frac{4m_{K}^2}{p^2} \Big)^{3/2}\Big]\nonumber\\
\Sigma_{K\pi}(p^2)&=&\frac{3F_0^2m_K^2}{16\pi^2
F_{K}^2F_{\pi}^2}\Big(1+\frac{3p^2}{4F_0^2}
  (g^2-\frac{g_A^4}{3\pi^2})\Big)\Big[
\frac{\lambda}{2}(1-\frac{p^2}{3m_K^2})+\int_{0}^{1}dx\Big(x
-x(1-x)\frac{p^2}{m_K^2}\Big)\ln\left|x-x(1-x)\frac{p^2}{m_{K}^2}\right|\nonumber\\
&&\;\;\;\;\;\;\;\;\;\;\;\;\;\;\;\;\;\;\;\;\;\;\;\;\;\;\;\;\;\;\;\;\;
 -\frac{i}{6}{\rm
Arg}(-1)\theta(p^2-(m_\pi+m_{K})^2)\frac{p^2}{m_K^2}\Big(
1-\frac{2(m_K^2+m_\pi^2)}{p^2}+\frac{(m_K-m_\pi)^4}{p^4}
\Big)^{3/2}\Big]
\end{eqnarray}
and
\begin{eqnarray}\label{6.16}
\rm Arg(-1)&=&(1+2k)\pi,\hspace{0.7in}k=0,\pm 1,\pm 2,\cdots,
\nonumber\\
\theta(x-y)&=&\left\{1,\hspace{1.1in}x>y; \atop
0,\hspace{1.1in}x\leq y.\right.
\end{eqnarray}
Note that $\Sigma_K(p^2)$ has no imaginary part for $E_{122}(p^2)$
and $E_{132}(p^2)$, while has imaginary part for $E_{33c}(p^2)$,
because $m_\rho,m_{K*}<2m_K$ while $m_\phi>2m_K$. As we can see,
two-point diagrams give an imaginary contribution, which results
from propagator of lighter pseudoscalar mesons. In
$\Sigma_\pi(p^2)$ and $\Sigma_{K\pi}(p^2)$, we have discarded
$m_\pi$ in the real part, leaving it in the imaginary part,
because it becomes important there. We also have neglected the
unimportant difference between $K$ and $\eta_8$ in $E_{132}(p^2)$.
There is no $\eta_8-\eta_8$ loop contribution in $E_{33c}$,
because it is constrained by space-like condition of external
vector mesons.

Considering all chain-like loop diagrams of complex pseudoscalar
loops, i.e., $\Sigma_\pi$, $\Sigma_{K\pi}$ and $\Sigma_K$ for
$\rho$, $K*$ and $\phi$ decay respectively, we have
\begin{eqnarray}\label{6.17}
S_{\rm chain}=\sum\limits_{abc}\int \frac{d^4p d^4q}{(2\pi2\pi)^4}
V_\mu^{ab}(p)\Phi^{bc}(-p-q)\Phi^{ca}(q)q^{\mu} E'_{abc}(p^2),
\end{eqnarray}
with
\begin{eqnarray}\label{6.18}
E'_{122}(p^2)&=&f^{(0)}_{111}(p^2)\frac{\Sigma_\pi(p^2)}{1-\Sigma_\pi(p^2)}+
f^{(0)}_{113}(p^2)\Sigma_K(p^2)\nonumber \\
E'_{132}(p^2)&=&f^{(0)}_{131}(p^2)\frac{\Sigma_{K\pi}(p^2)}{1-\Sigma_{K\pi}(p^2)}+
\frac{1}{2}\Big(f^{(0)}_{131}(p^2)+2f^{(0)}_{133}(p^2)\Big)\Sigma_K(p^2)
\nonumber \\
E'_{33c}(p^2)&=&3f^{(0)}_{331}(p^2)\frac{\Sigma_K(p^2)}{1-\Sigma_K(p^2)}
\end{eqnarray}

Similar consideration can also be applied to two-point diagram
correction to VMD vertex $\rho-\gamma$. The $\gamma-\Phi\Phi$
vertex (for quark loops see fig.2(a) and (b), with $V_\mu$
replaced by $A_\mu$) is
\begin{eqnarray}\label{6.19}
S_{\gamma-\Phi\Phi}=-e\sum\limits_{abc} \int \frac{d^4p
d^4q}{(2\pi2\pi)^4}
A_\mu(p)Q^{ab}\Phi^{bc}(-p-q)\Phi^{ca}(q)q^{\mu}
(\frac{F_0^2}{4}-f^{(0)}_{abc}).
\end{eqnarray}
Combining it with $S_{\rho-\Phi\Phi}$ (see eq.~(\ref{5.14})), we
obtain, after contractions and chain-like approximation,
pseudoscalar meson loop diagram correction (see fig.3(e)(f)) to
VMD vertex $\rho-\gamma$
\begin{eqnarray}\label{6.20}
S^{\rm chain}_{\rho-\gamma}=-\frac{e}{2} \int
\frac{d^4p}{(2\pi)^4} \rho_\mu^0(p)A^\mu(-p)E'_\gamma(p^2),
\end{eqnarray}
with
\begin{eqnarray}\label{6.21}
E'_\gamma(p^2)=-8\Big(\frac{f^{(0)}_{111}(p^2)}{F_\pi^2}\frac{\Sigma^\gamma_\pi(p^2)}{1-\Sigma_\pi(p^2)}
+\frac{f^{(0)}_{113}(p^2)}{F_K^2}\Sigma_K^\gamma(p^2)\Big)
\end{eqnarray}
where
\begin{eqnarray}\label{6.22}
\Sigma^\gamma_\pi(p^2)&=&-\frac{p^2(F_0^2-4f^{(0)}_{111}(p^2))}{4\pi^2F_\pi^2}
\Big[ \frac{\lambda}{6}
+\int_{0}^{1}dxx(1-x)\ln\frac{x(1-x)p^2}{m_{K}^2}+\frac{i}{6}{\rm
Arg}(-1)\theta(p^2-4m_\pi^2)\Big( 1-\frac{4m_\pi^2}{p^2}
\Big)^{3/2}\Big]\nonumber\\
\Sigma_K^\gamma(p^2)&=&\frac{m_K^2(F_0^2-4f^{(0)}_{113}(p^2))}{8\pi^2F_K^2}
\Big[
\lambda\Big(1-\frac{p^2}{6m_K^2}\Big)+\int_{0}^{1}dx\Big(1-x(1-x)\frac{p^2}{m_K^2}\Big)
\ln\Big(1-x(1-x)\frac{p^2}{m_{K}^2}\Big)\Big]
\end{eqnarray}

\section{Numerical Results}
\setcounter{equation}{0}

The widths for on-shell decays of vector mesons are determined by
\begin{eqnarray}\label{7.1}
\Gamma_{\rho \rightarrow e^+e^-}&=&\frac{\pi\alpha_{e.m.}^2}{3}
|\alpha(m_\rho^2)f_{\rho\gamma}^{(c)}(m_\rho^2)|^2m_\rho,
\nonumber\\
\Gamma_{\rho \rightarrow \pi\pi}&=&\frac{4}{3 \pi}
\Big|
\frac{\alpha(m_\rho^2)f_{111}^{(\rm c)}(m_\rho^2,m_{\pi}^2,m_{\pi}^2)
}{gF_{\pi}^2}
\Big|^2
m_\rho \Big[1-4\frac{m_{\pi}^2}{m_\rho^2}\Big]^{3/2},
\nonumber\\
\Gamma_{K^* \rightarrow K\pi}&=&\frac{1}{\pi}
\Big|
\frac{\alpha(m_{K^*}^2)
f_{132}^{(\rm c)}(m_{K^*}^2,m_{K}^2,m_{\pi}^2)
}{{\td g}F_{K}F_{\pi}}
\Big|^2
m_{K^*}\Big[1-2\frac{m_{K}^2+m_{\pi}^2}{m_{K^*}^2}+
\frac{(m_{K}^2-m_{\pi}^2)^2}{m_{K^*}^2}
\Big]^{3/2},
\nonumber\\
\Gamma_{\phi \rightarrow KK}&=&\frac{2}{3\pi}
\Big|
\frac{\alpha(m_{\phi}^2)
f_{331}^{(\rm c)}(m_{\phi}^2,m_{K}^2,m_{K}^2)
}{{\td g}F_{K}^2}
\Big|^2
m_{\phi}\Big[1-4\frac{m_{K}^2}{m_{\phi}^2}\Big]^{3/2}.
\end{eqnarray}
where $f_{\rho\gamma}^{(\rm c)}$ gets contributions from
eq.~(\ref{5.12}), (\ref{6.9}), and (\ref{6.20}), and
$f_{abc}^{(\rm c)}$ gets contributions from eq.~(\ref{5.10}),
(\ref{6.5}), and (\ref{6.13}):
\begin{eqnarray}\label{7.2}
f_{\rho\gamma}^{(\rm c)}(p^2)&=&f_{\rho\gamma}^{(0)}(p^2)
(1-\frac{e\lambda m_{K}^2}{8\pi^2F_{K}^2})+
E'_\gamma(p^2)/p^2, \nonumber \\
f_{abc}^{(\rm
c)}(p^2,q_1^2,q_2^2)&=&f^{(0)}_{abc}(p^2,q_1^2,q_2^2)+C_{abc}(p^2)+
E'_{abc}(p^2).
\end{eqnarray}
As indicated at the end of Sec.IV, we have included in eq.
~(\ref{7.1}) the rescaling factor $1/\td g \equiv
1/\sqrt{a_2-2f_1a_1}$ and the transformation factor
$\alpha(p^2)\equiv 1+f(p^2)p^2$ (which are both flavor-dependent)
for vector meson, and rescaling factor $2/F_\pi, 2/F_K$ for
external $\pi, K$ mesons respectively. In last equation for $\phi$
decay, we should distinguish $m_{K^\pm}$ from $m_{K^0}$ when we
consider $\phi \rightarrow K^+K^-$ and $\phi \rightarrow K_L^0
K_S^0$ respectively, because this is important for the difference
between the two decay widths.

Here are values of parameters: $m=460{\rm MeV}$ (chiral coupling
constant at $O(p^4)$, see Appendix A), $m_s=170{\rm MeV}$ (input),
$\kappa=0.5$ (chiral coupling constant at $O(p^4)$),
$\lambda=0.54$ (Zweig rule), $g=0.329$ (KSRF sum rule),
$g_{A}$=0.75 ($\beta$ decay of neutron), $F_0=F_\pi=185{\rm MeV}$,
$F_K=226{\rm MeV}$, $m_\pi=140{\rm MeV}$, $m_{K^\pm}=494{\rm
MeV}$, $m_{K^0}=497.6{\rm MeV}$, $m_\rho=769{\rm MeV}$,
$m_{K^*}=892{\rm MeV}$, $m_\phi=1019{\rm MeV}$. The numerical
results are listed in Table 1.

\begin{table}[hpb]
\centering
 \begin{tabular}{cccccc}
        & Leading Order
    & Resummation
    & non-chiral limit
    & After Loop Correction
    & Experimental Value \\
    &
    &
    & $f^{(0)}(m_V^2,m_{\Phi_1}^2,m_{\Phi_2}^2)$
    & $f^{(\rm c)}(m_V^2,m_{\Phi_1}^2,m_{\Phi_2}^2)$
    &
    \\ \hline
$\rho \rightarrow e^+e^-$&0.00465&0.00654& &0.00563&$0.00677 \pm 0.00032$ \\
$\rho \rightarrow \pi\pi$&123&187&194&175&$150.8\pm2.0$ \\
$K^* \rightarrow K\pi$&32.1&40.1&53.0&50.9&$50.7\pm0.9$ \\
$\phi \rightarrow K^+K^-$ &$^{\dag}0.951$&$^{\dag}1.494$&2.410&2.147&$2.193\pm0.031\pm0.016$ \\
$\phi \rightarrow K_L^0 K_S^0$ &$^{\dag}0.642$&$^{\dag}1.010$&1.643&1.465&$1.507\pm0.027\pm0.011$ \\
\end{tabular}
\begin{minipage}{5in}
\caption{Numerical results for vector mesons decays. These value
are in unit of MeV. The "leading order" and "resummation" columns
show results of momentum expansion obtained at chiral limit.
$^{\dag}$ These four values are listed just for uniformity and can
not be treated seriously, because $\phi$ physics should be studied
at non-chiral limit, which is required by unitarity. The two $f$s
in second line are not for $\rho \rightarrow e^+e^-$ decay.}
\end{minipage}
\end{table}

As we can see, the results after resummation differ large from the
leading-order ones, which shows that momentum expansion converges
slowly. Therefore, study at the leading order or the next order of
momentum expansion is very incomplete. Resummation is necessary
here. Moreover, loop corrections of pseudoscalar mesons, which are
next to the leading order of $N_c$ expansion, play an important
role, especially for SU(2) sector.

\section{unitarity and large $N_c$ expansion}
\setcounter{equation}{0}

Unitarity condition of $S$-matrix, or optical theorem,
\begin{eqnarray}\label{8.1}
{\rm Im}{\cal T}_{\beta,\alpha}=\frac{1}{2}\sum_{\rm all\;\gamma}
  {\cal T}_{\gamma,\alpha}{\cal T}_{\gamma,\beta}^*,
\end{eqnarray}
has to be satisfied for any well-defined quantum field theory,
where the ${\cal T}_{\beta,\alpha}$ is transition amplitude from
state $\alpha$ to state $\beta$, and $\gamma$ denotes all possible
intermediate states on mass shells. It is well-known that a low
energy effective meson theory should be a well-defined
perturbative theory in $N_c^{-1}$ expansion\cite{tH74}. Therefore,
we can expand ${\cal T}$-matrix in powers of $N_c^{-1}$,
\begin{eqnarray}\label{8.2}
{\cal T}=\sum_{n=0}^{\infty}{\cal T}_n,\hspace{1in} {\cal T}_n\sim
O((N_c^{-\frac{1}{2}})^n).
\end{eqnarray}
Then the unitarity condition of $S$-matrix for a low energy
effective meson theory has to satisfied order by order in powers
of $N_c^{-1}$,
\begin{eqnarray}\label{8.3}
{\rm Im}({\cal T}_{\beta,\alpha})_n=\frac{1}{2}\sum_{\rm
all\;\gamma(m\leq n)}({\cal T}_{\gamma,\alpha})_m
 ({\cal T}_{\gamma,\beta}^*)_{n-m}.
\end{eqnarray}

Now turn to the unitarity condition of $S$-matrix in the effective
meson field theory deduced from ChQM. First, from effective action
$S_n$ we can see that every vertex is of $O(N_c)$. From
eq.(\ref{4.4}) it can be showed that $g\sim O(\sqrt{N_c})$.
Moreover, eq.(\ref{3.5}) shows that there is a term
$(F_\pi^2/16)\int d^4x<\na_\mu U\na^\mu U^{\dag}>$ (after
renormalized) in $S_2$, which means that $F_\pi\sim
O(\sqrt{N_c})$. Because vector meson $V$ and pseudoscalar meson
$\Phi$ should be rescaled through $V\rightarrow V/g$ and
$\Phi\rightarrow\Phi/F_\pi$ respectively, there is no difference
between them as far as power counting about $N_c$ is concerned. In
what follows, the word "meson" means vector or pseudoscalar meson
if not indicated. So in any Feynman diagram every external meson
line is of $O(1/\sqrt{N_c})$ and every internal meson line is of
$O(1/N_c)$. Therefore, any transition amplitudes with $n_V$
vertex, $n_e$ external meson lines, $n_i$ internal meson lines and
$n_l$ loops of mesons are of order
\begin{eqnarray}\label{8.4}
N_c^{1-n_l-n_e/2}=(N_c^{-\frac{1}{2}})^{2n_l+n_e-2},
\end{eqnarray}
where relation $n_l=n_i-n_{_V}+1$ has been used.

Now consider transition amplitude from $n$ mesons state
$\{\alpha_1,\alpha_2,\cdots,\alpha_n\}$ to $k$ mesons state
$\{\beta_1,\beta_2,\cdots,\beta_k\}$. Assuming $\gamma$ is $s$
mesons state $\{\gamma_1,\gamma_2,\cdots,\gamma_s\}$, and using
the power counting rule (\ref{8.4}), eq.(\ref{8.3}) can be written
\begin{eqnarray}\label{8.5}
{\rm Im}({\cal
T}_{\beta,\alpha})_{(2n_l+k+n-2)}=\frac{1}{2}\sum_{{\rm
all}\;\gamma(s)}
 ({\cal T}_{\gamma,\alpha})_{(2n'_l+s+n-2)}
 ({\cal T}_{\gamma,\beta}^*)_{(2n''_l+s+k-2)},
\end{eqnarray}
where $n_l$, $n'_l$ and $n''_l$ are meson loop numbers of
transition amplitude ${\cal T}_{\beta,\alpha}$, ${\cal
T}_{\gamma,\alpha}$ and ${\cal T}_{\gamma,\beta}$ respectively.
Both side of eq. (\ref{8.5}) should be of the same order, thus
\begin{eqnarray}\label{8.6}
n'_l+n''_l+s=1+n_l.
\end{eqnarray}
At the leading order of transition $\alpha\rightarrow\beta$, we
have $n_l=0$, so $n'_l=n''_l=0$ and $s=1$. It means that when
summing over states $\gamma$ in eq. (\ref{8.5}), only one meson
state should be included.

What interests us is meson decay, i.e., $\alpha$ is one meson
state and $n=1$. Then $s=1$ indicates that only solution
$\gamma=\alpha$ is allowed at the leading order. However, for
$n'_l=0$, $({\cal T}_{\alpha,\alpha})_0\equiv0$, since meson
fields are free point-particle at limit
$N_c\rightarrow\infty$\cite{tH74}. Therefore, we have proved a
theorem that on-shell transition amplitude from one meson state to
any many mesons state must be real at leading order of $N_c^{-1}$
expansion,
\begin{eqnarray}\label{8.7}
{\rm Im}({\cal T}_{\beta,\alpha}^{(0)})_{(k-1)}=0,
\end{eqnarray}
where the superscript $(0)$ denotes leading order.

In the following we shall explicitly examine eq.~(\ref{8.1}) in
forward scattering of $\rho-$meson up to two-loop level of mesons.
The examination of other processes can be performed similarly. For
the case $\alpha=\beta=\rho$, $<\gamma|=<\pi\pi|$ is dominant.
Then for forward scattering of $\rho$-meson, eq.~(\ref{8.1})
becomes
\begin{eqnarray}\label{8.8}
\frac{2}{(2\pi)^4}{\rm Im}{\cal T}_{\rho,\rho}
 =\Gamma(\rho\rightarrow\pi\pi)=\frac{4}{3 \pi}
\Big[ \frac{\alpha(m_\rho^2)f_{111}^{(\rm
c)}(m_\rho^2,m_{\pi}^2,m_{\pi}^2) }{gF_{\pi}^2} \Big]^2 m_\rho
\Big[1-4\frac{m_{\pi}^2}{m_\rho^2}\Big]^{3/2},
\end{eqnarray}
and the expansions of ${\cal T}_{\rho,\rho}$ and $f_{111}^{(c)}$
are
\begin{eqnarray}\label{8.9}
{\cal T}_{\rho,\rho}=\sum_{n=0}^{\infty}({\cal
T}_{\rho,\rho})_{2n}, \hspace{1in}
f_{111}^{(c)}=\sum_{n=0}^{\infty}(f_{111}^{(c)})_{2n+1}.
\end{eqnarray}

At the leading order, Im$({\cal T}_{\rho,\rho})_0=0$ is obviously
satisfied. To obtain the imaginary part of $({\cal
T}_{\rho,\rho})_2$, we need to calculate meson loop correction in
fig. 5. The calculations are similar to those in Sect. VI, and the
result is
\begin{figure}[hptb]
\centerline{\psfig{figure=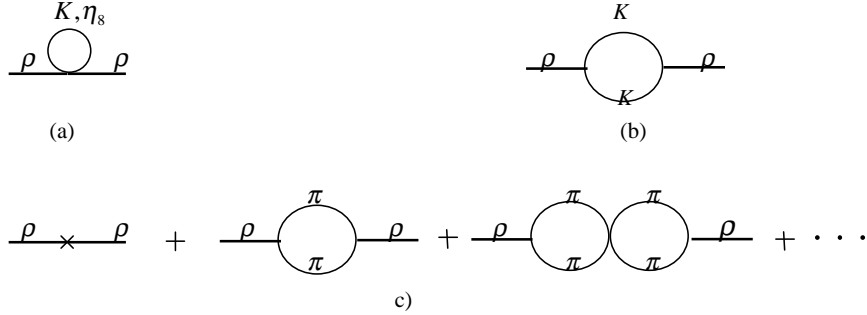,width=5in}} \centering
\begin{minipage}{5in}
   \caption{One-loop diagrams correcting to $\rho$ propagator.
    a)Tadpole diagram of $K$ or $\eta_8$. b)Two-point diagram of
    $K$. c)Chain-like approximation of pion.}
\end{minipage}
\end{figure}
\begin{eqnarray}\label{8.10}
{\cal L}_{\rho\rho}^{1-loop}=\frac{1}{2}\int\frac{d^4p}{(2\pi)^4}
 e^{-ip\cdot x}[\alpha(p^2)]^2\frac{64}{g^2}
 \Big\{\frac{[f_{111}^{(0)}(p^2)]^2}{F_\pi^4}\frac{\Sigma_\pi^{(0)}(p^2)}{1+\Sigma_\pi^{(0)}(p^2)}
   -\frac{[f_{113}^{(0)}(p^2)]^2}{F_K^4p^2}m_K^2\Sigma_K^{(0)}(p^2)\Big\}\nonumber\\
(g_{\mu\nu}p^2-p_\mu p_\nu)
  \rho_i^\mu(p)\rho_i^{\nu}(x),
\end{eqnarray}
where
\begin{eqnarray}\label{8.11}
\Sigma_\pi^{(0)}(p^2)&=&\frac{1}{8\pi^2}\{\frac{\lambda}{6}
  +\int_0^1dx\cdot x(1-x)\ln{\frac{x(1-x)p^2}{m_K^2}}
  +\frac{i}{6}Arg(-1)\theta(p^2-4m_\pi^2)
  (1-\frac{4m_\pi^2}{p^2})^{3/2}\},
  \nonumber \\
\Sigma_K^{(0)}(p^2)&=&\frac{1}{16\pi^2}\{\lambda(1-\frac{p^2}
  {6m_K^2})+\int_0^1dx(1-\frac{x(1-x)p^2}{m_K^2})
   \ln(1-\frac{x(1-x)p^2}{m_K^2})\}.
\end{eqnarray}
Taking $Arg(-1)=-\pi$, we obtain
\begin{eqnarray}\label{8.12}
\frac{2}{(2\pi)^4}{\rm Im}({\cal T}_{\rho,\rho})_2
 =\frac{4}{3 \pi}
\Big[
\frac{\alpha(m_\rho^2)f_{111}^{(0)}(m_\rho^2,m_{\pi}^2,m_{\pi}^2)
}{gF_{\pi}^2} \Big]^2 m_\rho
\Big[1-4\frac{m_{\pi}^2}{m_\rho^2}\Big]^{3/2}.
\end{eqnarray}
Noting that $(f_{111}^{(c)})_1=f_{111}^{(0)}$, we can see the
eq.~(\ref{8.8}) is satisfied at the next to leading order.

If the two-loop diagrams and three-loop diagrams are further
considered, we can prove the eq.~(\ref{8.8}) is satisfied up to
$O(N_c^{-3})$. The details of calculation are omitted here.

\section{determination of $\Lambda_{\rm CSSB}$}
\setcounter{equation}{0}

In the Section above, the unitarity of the effective meson theory
reduced from ChQM in terms of resummation calculations has been
investigated. By using optical theorem and large-$N_c$ expansion,
a theorem has been proved in framework of ChQM and at the leading
order of $N_c^{-1}$ expansion that the on-shell transition
amplitude ${\cal T}_{\beta,\alpha}^{(0)}$ (superscript (0) denotes
leading order of $N_c^{-1}$ expansion) must be real for transition
from one meson state $\alpha$ to any many mesons state $\beta$,
i.e.,
\begin{eqnarray}\label{9.1}
{\rm Im}({\cal T}_{\beta,\alpha}^{(0)})=0,
\end{eqnarray}
which, as we shall see, represents a nontrivial restriction on the
theory by the unitarity of $S$-matrix.

It is easy to see that, for $\alpha=\beta=V$ or $\Phi$, we have
\begin{eqnarray}\label{9.2}
{\rm Im}{\cal
T}_{V,V}^{(0)}=0,\;\;\;\;\;\;\;\;\;\;\;\;\;\;\;\;\;\;{\rm Im}{\cal
T}_{\Phi,\Phi}^{(0)}=0.
\end{eqnarray}
which satisfies the requirement of eq.~(\ref{9.1}). Next, we
examine ${\rm Im}{\cal T}_{\Phi\Phi, V}^{(0)}$. To calculate
transition amplitude from one vector meson state ($\alpha=V$) to
two pseudoscalar mesons state ($\beta=\{\Phi,\Phi$\}), we need
Feynman rules (see fig. 6) for vertex $V-\Phi\Phi$, which can be
read from action (\ref{5.10}) of the effective theory, and this is
at the leading order of $N_c^{-1}$ expansion.
\begin{figure}[hptb]
   \centerline{\psfig{figure=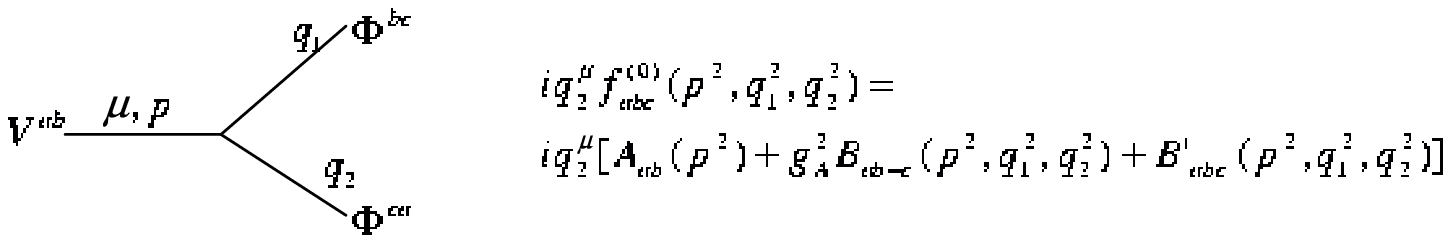,width=5.5in}}
 \centering
\begin{minipage}{5in}
   \caption{Feynman rule of vertex $V-\Phi\Phi$, where $A_{ab}(p^2)$,
   $B_{ab-c}(p^2,q_1^2,q_2^2)$ and $B'_{abc}(p^2,q_1^2,q_2^2)$
   are given by eqs. (\ref{5.2}), (\ref{5.5}) and (\ref{5.7})
   respectively.}
\end{minipage}
\end{figure}
We like to address that, being compared with ordinary quantum
field theories such as QED, QCD or $\phi^4$-theory, the vertex
function of our effective meson theory is much more complicated.
Especially, there are two objects in it (see eq. (\ref{5.2}),
(\ref{5.5}) and (\ref{5.7})),
$$
\ln\frac{D_1^{ab}}{m^2}=
\ln\frac{M_a^2(1-x)+M_b^2x}{m^2}+\ln\Big(1-\frac{p^2x(1-x)}{M_a^2(1-x)+M_b^2x}\Big),$$
$$\ln\frac{D_2^{abc}}{m^2}=\ln\frac{M_a^2x(1-y)+M_b^2(1-x)+M_c^2xy}{m^2}+\ln\Big(1-\frac{p^2x(1-x)(1-y)+q_1^2xy(1-x)
+q_2^2x^2y(1-y)}{M_a^2x(1-y)+M_b^2(1-x)+M_c^2xy}\Big).$$ There is
no such kind of logarithm terms in the Feynman rules of ordinary
fundamental quantum field theories, even in other kinds of
effective theories. This is a remarkably new feature. These two
logarithm terms result directly from resummation calculation and
they reflect non-perturbative effects, as should be emphasized.
There is no such kind of structure in any $O(p^4)$-, or
$O(p^6)$-effective meson theories.

Using the Feynman rule Fig.(6), we get transition amplitude for
$V\rightarrow 2\Phi$ at the leading order of $N_c^{-1}$ expansion
as follows
\begin{eqnarray}\label{9.3}
{\cal T}_{\Phi\Phi, V}^{(0)}\equiv
<\Phi^{bc}(q_1)\Phi^{ca}(q_2)|{\cal
T}^{(0)}|V^{ab}(p,\lambda)>=(2\pi)^4\delta^4(p-q_1-q_2)q_2^\mu
\epsilon_\mu^\lambda f^{(0)}_{abc}(p^2,q_1^2,q_2^2).
\end{eqnarray}
where $\epsilon_\mu^\lambda $ is the polarization vector of the
vector meson $V^{ab}(p,\lambda)$. Then, the imaginary part of
${\cal T}_{\Phi\Phi,V}^{(0)}$ reads
\begin{eqnarray}\label{9.4}
{\rm Im}{\cal T}_{\Phi\Phi,V}^{(0)}&=&\int_0^1 dx f(x,p){\rm
Im}\Big[\ln\Big(1-\frac{p^2x(1-x)}{M_a^2(1-x)+M_b^2x}\Big)\Big]
\nonumber
\\
&+&\int _0^1dx \int_0^1 dy g(x,y,p,q_1,q_2){\rm
Im}\Big[\ln\Big(1-\frac{p^2x(1-x)(1-y)+q_1^2xy(1-x)
+q_2^2x^2y(1-y)}{M_a^2x(1-y)+M_b^2(1-x)+M_c^2xy}\Big)\Big]
\end{eqnarray}
where $f(x,p)$ and $g(x,y,p,q_1,q_2)$ are definite real functions.
From theorem (\ref{9.1}), ${\rm Im}{\cal T}_{\Phi\Phi,V}^{(0)}$
has been required to be 0, otherwise, the unitarity of the theory
will be broken down. Consequently, we have
$$D_1^{ab}(p^2)\equiv M_a^2(1-x)+M_b^2x-p^2x(1-x)>0,\;\;\;\;\;\;(0\leq x\leq1),$$
and
$$
D_2^{abc}\equiv
M_a^2x(1-y)+M_b^2(1-x)+M_c^2xy-p^2x(1-x)(1-y)-q_1^2xy(1-x)
-q_2^2x^2y(1-y)>0,\;\;\;\;\;\;(0\leq x,y\leq1).
$$
This will lead to a restriction on the range of $p^2$. The former
inequality will hold in domain $0\leq x\leq1$ if
$$
M_{V^{ab}}=\sqrt{p^2}\leq M_a+M_b.
$$
As to the latter, the right side of it has no stationary point in
$x-y$ plane, therefore this inequality holding in the square
domain is equivalent to it holding at boundary of the square,
which gives
$$
\left\{\begin{array}{c}
 M_{V^{ab}}=\sqrt{p^2}\leq M_a+M_b,\\
 M_{\Phi^{ab}}=\sqrt{q^2}\leq M_a+M_b.
\end{array}
\right.
$$
Because $M_{\Phi^{ab}}<M_{V^{ab}}$, we see that the second
condition is satisfied if the first one does. Therefore we
conclude that the necessary condition for the effective theory to
be unitary is
\begin{eqnarray}\label{9.5}
M_{V^{ab}}=\sqrt{p^2}\leq 2m+m_a+m_b\equiv \Lambda^{ab} .
\end{eqnarray}
where $M_{a}=m+m_{a}\;\;(a=u,d,s)$ have been used.

Specifically, setting $(ab)=(ud)$ (i.e., $V^{ab}=\rho$,
$M_{V^{ab}}=m_\rho$) and $m_u=m_d=0$ in eq.(\ref{9.5}), we see
that, in ChQM with $\rho$ meson resonances, the unitarity of
$S$-matrix requires mass of constituent quark $m\geq
m_\rho/2$\cite{Wangvect}. In fact, this requirement is ensured in
ChQM by $m\simeq 460$MeV obtained by fitting the low energy limit
of the model (see Appendix A). For $V=K^*(892)$ and $\phi(1020)$
cases, $m_s\simeq 170MeV$ have to be taken into account, and the
corresponding unitarity conditions are $m\geq (m_{K^*}-m_s)/2$ and
$m\geq (m_{\phi}-2m_s)/2$ respectively. They are also satisfied by
taking $m\simeq 460$MeV.

In the above discussions, we have actually revealed an important
fact that $\Lambda^{ab} \equiv 2m+m_a+m_b$ is a critical energy
scale in the effective meson theory of ChQM. As $\sqrt{p^2}$ is
below $ \Lambda^{ab}$, the $S$-matrices yielded from the Feynman
rules of the meson theory are unitary, while as $\sqrt{p^2}$ is
above this scale, the unitarity of the meson theory will be broken
down. This fact indicates that the well-defined effective quantum
field theory describing the meson physics in the framework of ChQM
exists only as the typical energies are below $\Lambda^{ab}$. When
energy is above $\Lambda^{ab}$, the effective meson-Lagrangian
description of the dynamics is illegal in principle because the
unitarity fails. This is precisely a critical phenomenon, or
quantum phase transition in quantum field theory, which is caused
by quantum fluctuations in the system\cite{Sachdev}. Recalling the
meaning of the scale $\Lambda_{\rm CSSB}$ of chiral symmetry
spontaneously broking in QCD, we can see that $\Lambda^{ab}$ play
the same role as $\Lambda_{\rm CSSB}$. Then, in the framework of
ChQM we have
\begin{equation}\label{9.6}
\Lambda_{\rm CSSB}=\Lambda^{ab}\equiv 2m+m_a+m_b.
\end{equation}
It is essential here that $\Lambda_{\rm CSSB}$ is
flavor-dependent. Numerically, for $ud$-flavor system
(e.g.,$\pi-\rho-\omega$ physics),
\begin{equation}\label{ud}
\Lambda_{\rm CSSB}(ud)\simeq 2m=920MeV, \end{equation}
 for$u(d)s$-flavor system (e.g., $K-K^*$ physics),
\begin{equation}\label{ud-s}
\Lambda_{\rm CSSB}(u(d)s)\simeq 2m+m_s=1090MeV,
\end{equation}
 for $\bar{s}s$ case
(e.g., $\phi$-physics),
\begin{equation}\label{ss}
\Lambda_{\rm CSSB}(ss)\simeq 2(m+m_s)=1260MeV.
\end{equation}
Since $m_\rho<\Lambda_{\rm CSSB}(ud)$,$m_{K^*}<\Lambda_{\rm
CSSB}(u(d)s)$ and $m_\phi<\Lambda_{\rm CSSB}(ss)$, the effective
meson field theory derived by resummation derivation in ChQM in
this paper is unitary. And the low energy expansions of $p$ are
legitimate and convergent due to $p^2/\Lambda_{\rm CSSB}^2<1$. It
means that all light flavor vector meson resonances can be
included in ChQM model consistently. It is remarkable that the
quantum phase transitions in ChQM can be explored successfully in
resummation derivation method, and the corresponding critical
scales are determined analytically.

In ref.\cite{MG84}, $\Lambda_{\rm CSSB}$ has been estimated by
comparing ${\cal O}(p^2)$ contributions to ${\cal O}(p^4)$'s for
$\pi-\pi$ scattering process, and it has been shown that
$\Lambda_{\rm CSSB}$ is around $2\pi F_\pi\sim 1.2GeV$. However,
the quantum phase transitions in ChQM were not explored in
ref.\cite{MG84}, and the existence of $\Lambda_{\rm CSSB}$ is a
prior hypothesis there. Therefore the studies on determination of
$\Lambda_{\rm CSSB}$ in this present paper is significantly
different from ones in \cite{MG84}, even though the numerical
values of $\Lambda_{\rm CSSB}$ both in \cite{MG84} and in this
paper are closing.

\section{Summary and Discussion}
\setcounter{equation}{0}

In this paper, starting from the chiral constituent quark model
with the lowest vector meson resonances we have achieved
resummation studies on the processes of $\rho \rightarrow \pi\pi$,
$\rho^0 \rightarrow e^+e^-$, $K^* \rightarrow K\pi$, $\phi
\rightarrow K^+K^-$ and $\phi \rightarrow K^0_LK^0_S$, and the
results are in good agreement with the data. The error is less
than $17\%$. In our calculations, all chiral expansion (or
$p$-expansion) terms have been included, and analytic expressions
for decay widthes of these processes are obtained. Distinguishing
from the ChQM-effective Lagrangian derivations existed in the
literature\cite{Chan85,ENJL,Li95}, we have calculated not only the
quark loops, but also the Goldstone boson loops at one-loop level.
Thus, besides the contributions of leading order of
$1/N_c$-expansion, ones due to the next to the leading order of
$1/N_c$ have also been taken into account. The logarithmic
divergence due to quark loops and the quadratic divergence due to
meson loops have been absorbed by the parameters $g$ and $\lambda$
respectively, while the values of $g$ and $\lambda$ are determined
by the KSRF sum rule and by the Zweig rule forbidden to
$\phi\rightarrow \pi\pi$ respectively. The fact that this
QCD-inspired parametrization leads to the reasonable results
indicates that both the model employed by us and the derivations
in this paper are legitimate and consistent.

In the Introduction, we have addressed the resummation studies are
necessary for the vector meson resonance physics because it is
related to the question whether the chiral dynamical description
is legitimate or not. The success of such studies achieved in this
paper provides a meaningful evidence that the chiral Lagrangian
method wokes at this energy region. This is one of the main
conclusions of this paper. It should be addressed again that this
conclusion can not be reached from any so called $O(p^4)$- or
$O(p^6)$-studies on any chiral Lagrangian theories or models with
vector meson resonances. Actually, any predictions coming from a
chiral Lagrangian with only few terms in chiral expansion in this
energy region are not reliable. In the TABLE I, we have shown that
the contributions coming from high-order terms of $p$-expansion
are important. It implies that the feasibility to evaluate high
order contribution of momentum expansion has to be ensured for any
practically working effective meson theories with vector meson
resonances. In addition we have also point out in the Introduction
that the contributions of the next to leading order in
$1/N_c$-expansion have also to be considered, otherwise the loop
calculations will be incomplete. Our calculations show that the
situation is as expected indeed. The next to leading order
contributions in the expansion of $1/N_c$ makes the predictions
more closing to the data significantly (see TABLE I).

The unitarity of the effective meson theory including all
$p$-expansion terms deduced from ChQM has been investigated by
means of the optic theorem and $1/N_c$ expansion argument in QCD.
It has been found that the necessary condition for the unitarity
of the theory is the momentum $p$ of vector meson $V^{ab}$
satisfies $\sqrt{p^2}< \Lambda^{ab}\equiv 2m+m_a+m_b$, otherwise,
as $\sqrt{p^2}$ is above $\Lambda^{ab}$, the unitarity will be
broken down. Then, we conclude that the chiral symmetry
spontaneously breaking scale $\Lambda_{\rm CSSB}=\Lambda^{ab}$. It
is clear that this represents an explicit study on a quantum
critical phenomenon. Actually, we are working in order phase (or
meson field phase), i.e., the order parameter is constituent quark
mass $m\neq 0$, dynamical quantum field freedoms are meson fields
and the quantum fluctuation parameter is ${\cal G
}=p^2/(\Lambda^{ab})^2$. And then we revealed that the critic
point is ${\cal G}_c=1$. Obviously, our method to study CSSB is
significantly different from ones in\cite{Nambu}\cite{Roberts}
where the gap equation (or Schwinger-Dyson equation) is used for
understanding CSSB.

An important feature of this result is that $\Lambda_{\rm CSSB}$
in ChQM is flavor-dependent, i.e., $\Lambda_{\rm CSSB}(ud)\neq
\Lambda_{\rm CSSB}((ud)s)\neq \Lambda_{\rm CSSB}(ss)$. This result
should be interesting and meaningful in physics because the
critical energy scale is one of quantities to characterize the
intrinsic properties of the physical system, and hence it should
be dependent of the contents on the system. Furthermore, that
$m_\rho< \Lambda_{\rm CSSB}(ud)$, $m_{K^*}< \Lambda_{\rm
CSSB}((ud)s)$ and $m_\phi< \Lambda_{\rm CSSB}(ss)$ indicates that
the  studies on the vector meson decays presented in this paper in
the chiral effective meson field theory of ChQM are legitimate and
self-consistent.

\begin{center}
{\bf ACKNOWLEDGMENTS}
\end{center}
This work is partially supported by NSF of China 90103002.

\appendix

\section{Low energy limit}

It is well known that, at very low energy, ChPT is a rigorous
consequence of the symmetry pattern of QCD and its spontaneous
breaking. Therefore the low energy limit of any models concerning
meson resonances must match with ChPT. The low energy limit of
ChQM model can be obtained via integrating over vector meson
resonances. It means that, at very low energy, the dynamics of
vector mesons are replaced by pseudoscalar meson fields. Since
there are no interaction of vector mesons at $O(p^2)$, at very low
energy, the equation of motion $\delta{\cal L}/\delta V_\mu=0$
yields classical solution for vector mesons
\begin{equation}\label{A1.1}
V_\mu=\frac{1}{m_{_V}^2}\times {\rm terms\;of\;} O(p^3),
\end{equation}
where $p$ is momentum of pseudoscalar at very low energy.
Therefore, in effective action $S_n$, the terms involving vector
meson resonances are $O(p^6)$ at very low energy and do not
contribute to $O(p^4)$ low energy coupling constants,
$L_i\;(i=1,2,...,10)$. The low energy coupling constants
$L_i\;(i\neq 7)$ yielded by ChQM model can be directly obtained
from effective actions resulted from quark loop,
\begin{eqnarray}\label{A1.2}
L_1&=&\frac{1}{2}L_2=\frac{1}{128\pi^2}, \hspace{0.8in}
L_3=-\frac{3}{64\pi^2}+\frac{1}{64\pi^2}g_A^4, \nonumber \\
L_4&=&L_6=0, \hspace{1.2in}L_5=\frac{3m}{32\pi^2B_0}g_A^2,\nonumber \\
L_8&=&\frac{F_0^2}{128B_0m}(3-\kappa^2)+\frac{3m}{64\pi^2B_0}
  (\frac{m}{B_0}-\kappa g_A-\frac{g_A^2}{2}-\frac{B_0}{6m}g_A^2)
  +\frac{L_5}{2},\nonumber \\
L_9&=&\frac{1}{16\pi^2}, \hspace{1.35in}
L_{10}=-\frac{1}{16\pi^2}+\frac{1}{32\pi^2}g_A^2.
\end{eqnarray}
In fact, the above expression on $L_i$ have been obtained in some
previous refs.\cite{MG84,Esp90,Bijnens93} (besides of $L_8$).

The constants $L_7$ has been known to get dominant contribution
from $\eta_0$\cite{GL85a} and this contribution is suppressed by
$1/N_c$. If we ignore the $\eta-\eta^{\prime}$ mixing, we have
\begin{equation}\label{A1.3}
  L_7=-\frac{f_\pi^2}{128m_{\eta^{\prime}}^2}.
\end{equation}

Thus six free parameters, $g$ (fitted by KSRF sum rule), $g_{_A}$
(fitted by $n\rightarrow pe^-\bar{\nu}_e$ decay), $B_0$, $\kappa$,
$m$ and $m_{\eta^{\pr}}$ determine all ten low energy coupling
constants of ChPT. It reflects the dynamics constrains between
those low energy coupling constants. Moreover, if we take
$m_u+m_d\simeq 11$MeV, we can obtain
$B_0=\frac{m_\pi^2}{m_u+m_d}\simeq 1.8$GeV. Then experimental
values of $L_5$ constrains constituent quark mass $m\simeq
460$MeV. Setting $F_0=F_\pi$ and using $L_8$ as input, we find
that $\kappa\simeq0.5$. The numerical results for these low energy
constants are in table II.

\begin{table}[pb]
\centering
 \begin{tabular}{ccccccccccc}
&$L_1$&$L_2$&$L_3$&$L_4$&$L_5$&$L_6$&$L_7$&$L_8$&$L_9$&$L_{10}$
  \\ \hline
ChPT&$0.7\pm 0.3$&$1.3\pm 0.7$&$-4.4\pm 2.5$&$-0.3\pm 0.5$&$1.4\pm
  0.5$&$-0.2\pm 0.3$&$-0.4\pm 0.15$
  &$0.9\pm 0.3$&$6.9\pm 0.7$&$-5.2\pm 0.3$\\
{ChQM}&0.79&1.58&-4.25&0&$1.4^{a)}$&0&$(-0.4\pm 0.1)^{b)}$&
 $0.9^{a)}$&6.33&-4.55
   \end{tabular}
\begin{minipage}{6in}
\caption{\small $L_i$ in units of $10^{-3}$, ${\mu}=m_\rho$.
a)input. b)contribution from gluon anomaly.}
\end{minipage}
\end{table}

\section{cancellation of quadratic divergence of meson loops}

From calculations in Sect. VI, we can find that only quadratic
divergence appears in one-loop contribution of pseudoscalar
mesons. Since the present model is a non-renormalizable effective
theory, the divergences have to be factorized, i.e., the parameter
$\lambda$ has to be determined phenomenologically.

The on-shell decay $\phi\rightarrow\pi\pi$ is forbidden by $G$
parity conservation and Zweig rule. Experiment also show that
branching ratios of this decay is very small,
$B(\phi\rightarrow\pi\pi)=(8\;{+5\atop -4})\times 10^{-5}$.
Theoretically, this decay can occur through photon-exchange or
$K$-loop (fig.5). The latter two diagrams yield non-zero imaginary
part of decay amplitude. Thus the real part yielded by the latter
two diagrams must be very small. We can determine $\lambda$ due to
this requirement. From the calculation in the above two
subsection, we see that result yielded by the latter two diagrams
is proportional to a factor
\begin{equation}\label{B2.1}
\lambda(\frac{p^2}{6}-m_{_K}^2)-\int_0^1dx\cdot
[m_{_K}^2-x(1-x)p^2]
   \ln{(1-\frac{x(1-x)p^2}{m_{_K}^2})}.
\end{equation}
\begin{figure}[hptb]
\centerline{\psfig{figure=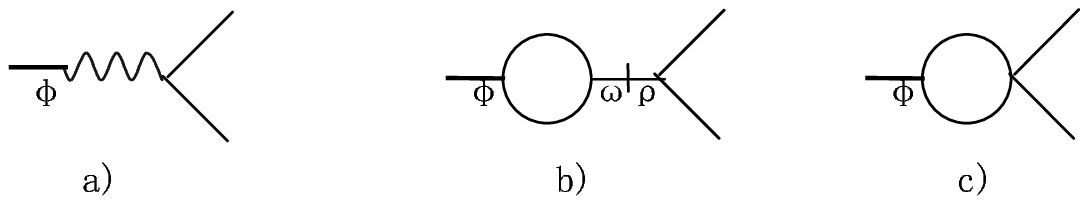,width=5in}} \centering
\begin{minipage}{5in}
   \caption{Some diagrams for $\phi\rightarrow\pi\pi$ decay. The
   one-loop in figure b) and c) is $K$-loop.}
\end{minipage}
\end{figure}
Then Zweig rule requires
\begin{equation}\label{B2.2}
\Big \{\lambda(\frac{p^2}{6}-m_{_K}^2)-{\rm Re}\int_0^1dx\cdot
 [m_{_K}^2-x(1-x)p^2]\ln{(1-\frac{x(1-x)p^2}{m_{_K}^2})}\Big \}
 \Big |_{p^2=m_\phi^2}\simeq 0.
\end{equation}
Form the above equation, we obtain $\lambda\simeq 0.54$.

\end{document}